\documentclass[]{spie} 

\newcommand{\Spider}{\textsc{Spider}}

\usepackage[]{graphicx}
\usepackage{amssymb}
\usepackage{amsmath}
\usepackage[T1]{fontenc}
\usepackage[usenames,dvipsnames,svgnames,table]{xcolor}
\usepackage{indentfirst}
\usepackage[colorlinks=true]{hyperref}
\hypersetup{
	linkcolor=RoyalBlue,
	citecolor=RoyalBlue,
	filecolor=RoyalBlue,
	urlcolor=RoyalBlue,
}

\title{Pointing control for the {\LARGE \textsc{\bfseries Spider}} balloon-borne telescope} 

\author{J.\,A.~Shariff\supit{a}$^\dagger$, P.\,A.\,R.~Ade\supit{b}, M.~Amiri\supit{c}, S.\,J.~Benton\supit{d}, J.\,J.~Bock\supit{e}\supit{,f}, J.\,R.~Bond\supit{g}\supit{,h},\\S.\,A.~Bryan\supit{i}, H.\,C.~Chiang\supit{j}, C.\,R.~Contaldi\supit{k}, B.\,P.~Crill\supit{e}\supit{,f}, O.\,P.~Dor\'e\supit{e}\supit{,f}, M.~Farhang\supit{a}\supit{,g}, J.\,P.~Filippini\supit{e}, L.\,M.~Fissel\supit{a}\supit{,l}, A.\,A.~Fraisse\supit{m}, A.\,E.~Gambrel\supit{m}, N.\,N.~Gandilo\supit{a}, S.\,R.~Golwala\supit{e}, J.\,E.~Gudmundsson\supit{m}, M.~Halpern\supit{c}\supit{,h}, M.~Hasselfield\supit{n}, G.\,C.~Hilton\supit{o}, W.\,A.~Holmes\supit{f}, V.\,V.~Hristov\supit{e}, K.\,D.~Irwin\supit{o}\supit{,p}\supit{,q}, W.\,C.~Jones\supit{m}, Z.\,D.~Kermish\supit{m}, C.\,L.~Kuo\supit{p}\supit{,q}, C.\,J.~MacTavish\supit{r}, P.\,V.~Mason\supit{e}, K.\,G.~Megerian\supit{f}, L.~Moncelsi\supit{e}, T.\,A.~Morford\supit{e}, J.\,M.~Nagy\supit{i},\\C.\,B.~Netterfield\supit{a}\supit{,d}\supit{,h}, R.~O'Brient\supit{e,f}, A.\,S.~Rahlin\supit{m}, C.\,D.~Reintsema\supit{o}, J.\,E.~Ruhl\supit{i}, M.\,C.~Runyan\supit{f}, J.\,D.~Soler\supit{a}\supit{,s}, A.~Trangsrud\supit{f}, C.\,E.~Tucker\supit{b}, R.\,S.~Tucker\supit{e}, A.\,D.~Turner\supit{f}, A.\,C.~Weber\supit{f}, D.\,V.~Wiebe\supit{c}, and E.\,Y.~Young\supit{m}
\skiplinehalf
\supit{a}Department of Astronomy \& Astrophysics, University of Toronto, Toronto, ON, Canada;\\
\supit{b}School of Physics and Astronomy, Cardiff University, Cardiff, UK;\\
\supit{c}Department of Physics \& Astronomy, University of British Columbia,\\Vancouver, BC, Canada;\\
\supit{d}Department of Physics, University of Toronto, Toronto, ON, Canada;\\
\supit{e}Division of Physics, Mathematics \& Astronomy, California Institute of Technology,\\Pasadena, CA, USA;\\
\supit{f}Jet Propulsion Laboratory, Pasadena, CA, USA;\\
\supit{g}Canadian Institute for Theoretical Astrophysics, Toronto, ON, Canada;\\
\supit{h}Canadian Institute for Advanced Research, Toronto, ON, Canada;\\
\supit{i}Department of Physics, Case Western Reserve University, Cleveland, OH, USA;\\
\supit{j}School of Mathematics, Statistics \& Computer Science, University of KwaZulu-Natal,\\Durban, South Africa;\\
\supit{k}Theoretical Physics, Blackett Laboratory, Imperial College, London, UK;\\
\supit{l}CIERA, Northwestern University, Evanston, IL, USA;\\
\supit{m}Department of Physics, Princeton University, Princeton, NJ, USA;\\
\supit{n}Department of Astrophysical Sciences, Princeton University, Princeton, NJ, USA;\\
\supit{o}National Institute of Standards and Technology, Boulder, CO, USA;\\
\supit{p}Department of Physics, Stanford University, Stanford, CA, USA;\\
\supit{q}Kavli Institute for Particle Astrophysics and Cosmology,\\SLAC National Accelerator Laboratory, Menlo Park, CA, USA;\\
\supit{r}Kavli Institute for Cosmology, University of Cambridge, Cambridge, UK;\\
\supit{s}Institut d'Astrophysique Spatiale, CNRS \& Universit\'e Paris-Sud, Orsay, France;\\
}

\authorinfo{\noindent $^\dagger$Corresponding author; E-mail: shariff@astro.utoronto.ca, Tel: 1~416~946~0946\\Copyright 2014 Society of Photo-Optical Instrumentation Engineers. One print or electronic copy may be made for personal use only. Systematic reproduction and distribution, duplication of any material in this paper for a fee or for commercial purposes, or modification of the content of the paper are prohibited.}

\begin{document} 
\maketitle 

\begin{abstract}
We present the technology and control methods developed for the pointing system of the \textsc{Spider} experiment. \textsc{Spider} is a balloon-borne polarimeter designed to detect the imprint of primordial gravitational waves in the polarization of the Cosmic Microwave Background radiation. We describe the two main components of the telescope's azimuth drive: the reaction wheel and the motorized pivot. A 13~kHz PI control loop runs on a digital signal processor, with feedback from fibre optic rate gyroscopes. This system can control azimuthal speed with $<$ 0.02~deg/s RMS error. To control elevation, \textsc{Spider} uses stepper-motor-driven linear actuators to rotate the cryostat, which houses the optical instruments, relative to the outer frame. With the velocity in each axis controlled in this way, higher-level control loops on the onboard flight computers can implement the pointing and scanning observation modes required for the experiment. We have accomplished the non-trivial task of scanning a 5000~lb payload sinusoidally in azimuth at a peak acceleration of 0.8~deg/s$^2$, and a peak speed of 6~deg/s. We can do so while reliably achieving sub-arcminute pointing control accuracy.
\end{abstract}



\section{INTRODUCTION}
\label{sec:intro}  

\Spider~is an experiment designed to detect the curl-like or ``$B$-mode'' component of the Cosmic Microwave Background (CMB) polarization. This component, if present, would be an imprint of the primordial gravitational waves predicted to have been produced by inflation in the very early universe. The amplitude of the gravitational waves is parameterized by the tensor-to-scalar ratio, $r$, which is the ratio of the amplitude of the primordial power spectrum of tensor perturbations to that of scalar perturbations at a particular spatial scale. A measurement of $r$ would set the energy scale at which inflation occurred, providing a much needed constraint on theoretical models of inflation. Detecting the faint signature of $B$-mode polarization requires an experiment with unprecedented sensitivity and control over systematics. \Spider's approach to this problem, in part, is to observe from Antarctica at 36 km altitude, aboard a stratospheric balloon. During the austral summer (December to January), circumpolar winds enable a 20-day Long Duration Balloon (LDB) flight around the continent, which provides the necessary integration time and sky coverage.

\Spider~consists of six monochromatic, refracting, millimetre-wave telescopes (or ``receivers'')~\cite{runyan2010}, all enclosed within and cooled by a 2 m diameter, 1284~L liquid helium cryostat~\cite{gudmunds2010}. At the focal plane of each receiver is a large-format array of spatial pixels, each containing two detectors with orthogonal polarization sensitivity. These detectors are antenna-coupled, superconducting Transition Edge Sensor (TES) bolometers~\cite{kuo2008}. They are read out by time domain-multiplexed current amplifiers known as Superconducting Quantum Interference Devices (SQUIDs). Three 90~GHz receivers and three 150~GHz receivers are planned for the initial \Spider~flight configuration, for a total of 2400 bolometers. A rotating, stepped half-wave plate (HWP)~\cite{bryan2010}, located skyward of the primary optic, modulates the polarization of the sky signal, allowing each detector to make an independent measurement of the $Q$ and $U$ Stokes parameters for linear polarization. Combined with a fast scan strategy, this mitigates systematic effects that mix CMB temperature to polarization. \Spider~was designed to be able to set an upper limit on the tensor-to-scalar ratio of $r<0.03$ at 3$\sigma$~\cite{jpf2010}. A general overview of the \Spider~experiment, including more details of its scientific goals and pre-flight integration status, is given elsewhere in these proceedings~\cite{asr2014}.

Observations from a stratospheric balloon are enabled by \Spider's flight systems, which are located on the gondola. The gondola and its flight systems were designed, assembled, and tested in the Balloon Astrophysics Group at the University of Toronto. The gondola is a strong and lightweight support structure for the flight cryostat~\cite{jds2014}. It is also a fully-automated, pointable platform for the experiment. The gondola includes two redundant onboard flight computers, which control the experiment's operation during flight. Another key component of the gondola is the Attitude Control System (ACS). The ACS includes BLASTbus electronics~\cite{sjb2014}, which provide general purpose readout and control for the gondola's various sensors, actuators, and other subsystems. The gondola carries a suite of pointing sensors: star cameras, a differential GPS, a magnetometer, inclinometers, pinhole sun sensors, rate gyroscopes, and rotary encoders. These components are used for in-flight and post-flight reconstruction of the telescope pointing~\cite{nng2014}. Note that Refs.~\citenum{jds2014},~\citenum{sjb2014}, and~\citenum{nng2014} are also submissions to these proceedings.

\begin{figure}[hbtp!]
\begin{center}
\includegraphics[width=0.47\textwidth]{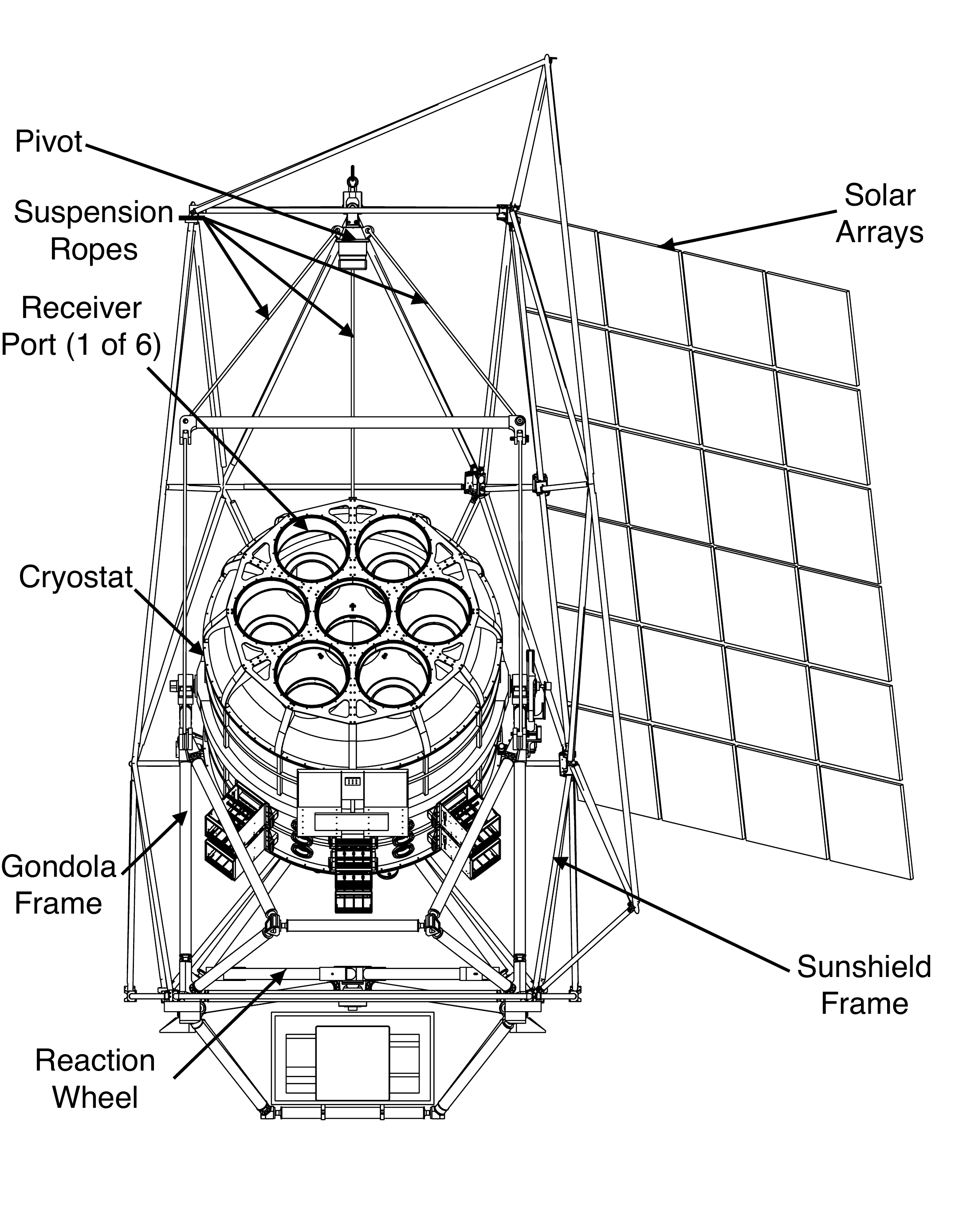}
\includegraphics[width=0.40\textwidth]{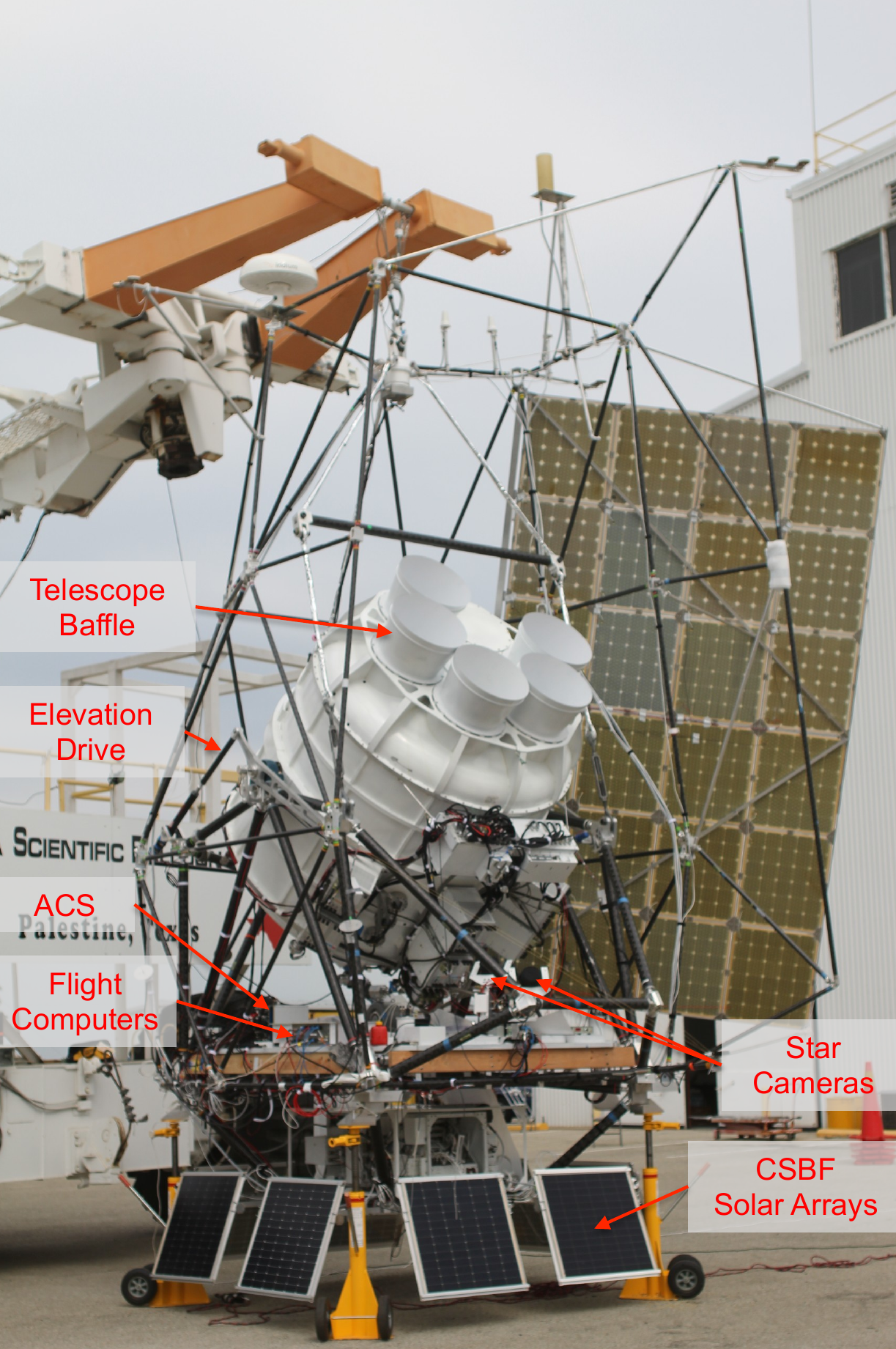}
\caption{\label{fig:spider} \textit{Left} -- A drawing of the \Spider~payload with major structural elements and components of the pointing system labelled. \textit{Right} -- A photograph of the fully-assembled \Spider~payload hanging from the launch vehicle at the NASA Columbia Scientific Balloon Facility (CSBF) in Palestine, Texas. Indicated on the photo are some features not displayed in the drawing, including telescope baffles, the elevation drive, the CSBF solar panels at the bottom of the payload, and the main deck to which all of the flight electronics are mounted.}
\end{center}
\end{figure}

The fully-assembled \Spider~payload includes the cryostat with receivers, the gondola and flight systems, the sunshield framework, and the solar arrays. Figure~\ref{fig:spider} shows a drawing and a photograph of the payload with key features labelled. Control of a payload's attitude while it is hanging from the balloon is a key problem for any balloon-borne astrophysical experiment. The solutions to this problem implemented on \Spider~build from those developed for past such experiments, including \textsc{BOOMERanG}~\cite{crill2003}, BLAST~\cite{pascale2008}, and BLASTPol~\cite{fea2014}, but with several important innovations that are described herein.

This paper is divided into five sections. Pointing control in azimuth and in elevation is discussed in Sections \ref{sec:az} and \ref{sec:el} respectively. Section \ref{sec:obs} explains how this low-level control of \Spider's motion in two orthogonal axes is used to implement higher-level scan modes. This section includes an overview of the main observing strategy planned for \Spider's first flight. Finally, Section \ref{sec:sys} evaluates the performance of the pointing control system using data from \Spider~scan tests.

\section{THE AZIMUTH DRIVE}
\label{sec:az}
The \Spider~azimuth (az) drive has two primary components: the reaction wheel and the motorized pivot. The reaction wheel is supported by four carbon fibre struts extending upward and inward from the bottom corners of the gondola frame (Figures~\ref{fig:spider} and~\ref{fig:wheel}). It sits just below the main aluminum honeycomb floor of the gondola, where the electronics are mounted. The reaction wheel works by conservation of angular momentum: a torque applied to the reaction wheel by its motor results in a torque of opposite sign on the gondola. The pivot is a motorized joint located at the top of the payload, where the three suspension cables meet (Figure~\ref{fig:spider}). Extending upwards from the pivot rotor is the flight train: the steel cabling that connects to the balloon. The pivot stator is rigidly connected to the suspension cables and the rest of the payload underneath. The pivot is able to provide additional torque in az by twisting the flight train, which acts as a torsional spring. This torque aids in scanning the gondola, and prevents the reaction wheel from reaching its saturation speed: the speed at which the motor's back-EMF\footnote{For these motors, the back-EMF $V_b = K_b\omega$, where $K_b$ is the voltage constant (Table~\ref{tab:motors})} prevents further current through the motor windings. Without the pivot, saturation inevitably occurs because the reaction wheel must absorb the angular momentum produced by external torques caused by wind shear and other disturbances. On long timescales, the pivot is able to dump this excess angular momentum to the balloon, through the flight train.  

\subsection{The Reaction Wheel}
\label{sec:rw}

\subsubsection{Mechanical Design}
\label{sec:rwstruct}

\begin{figure}[htbp!]
\includegraphics[width=0.5\textwidth]{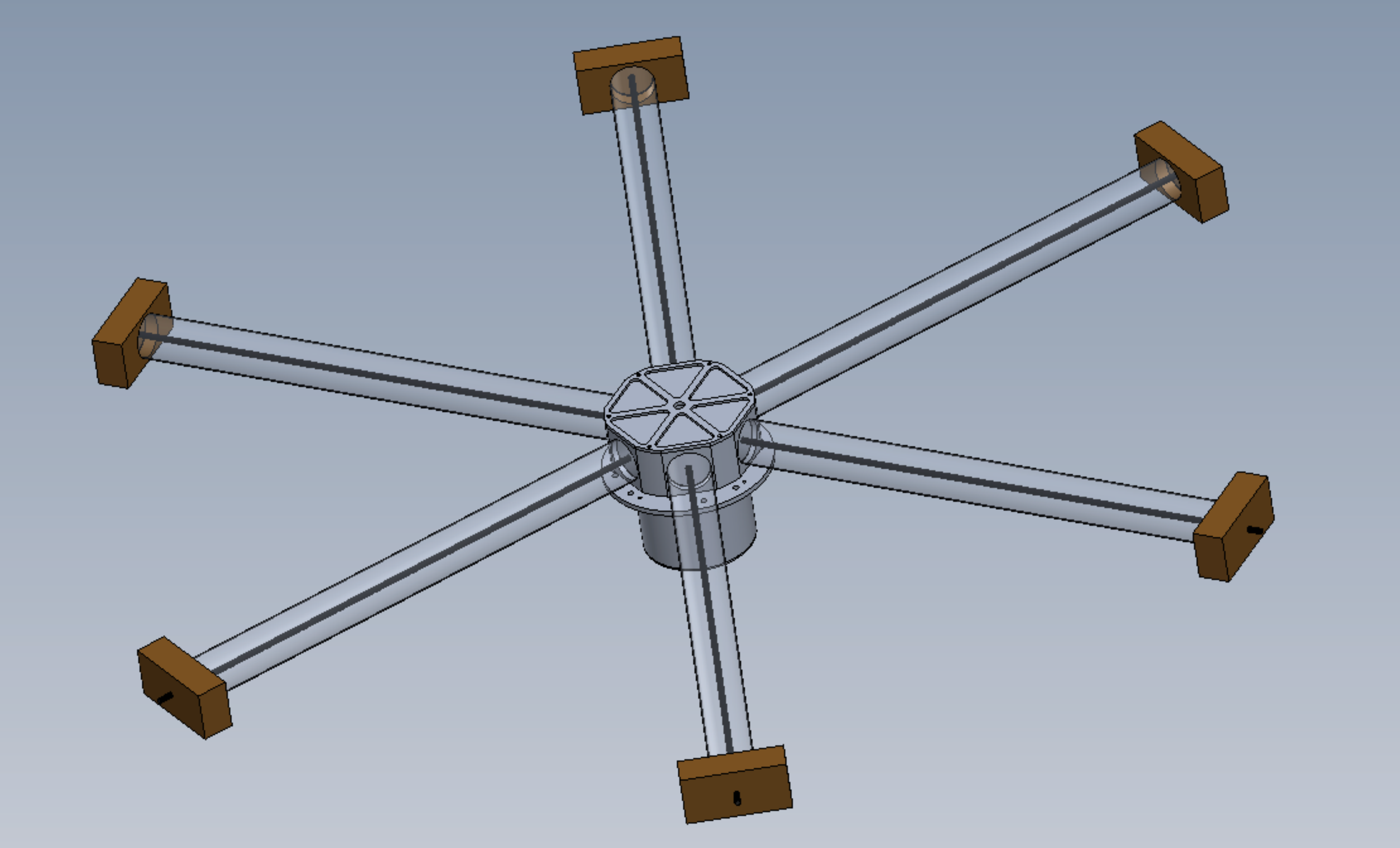}
\includegraphics[width=0.5\textwidth]{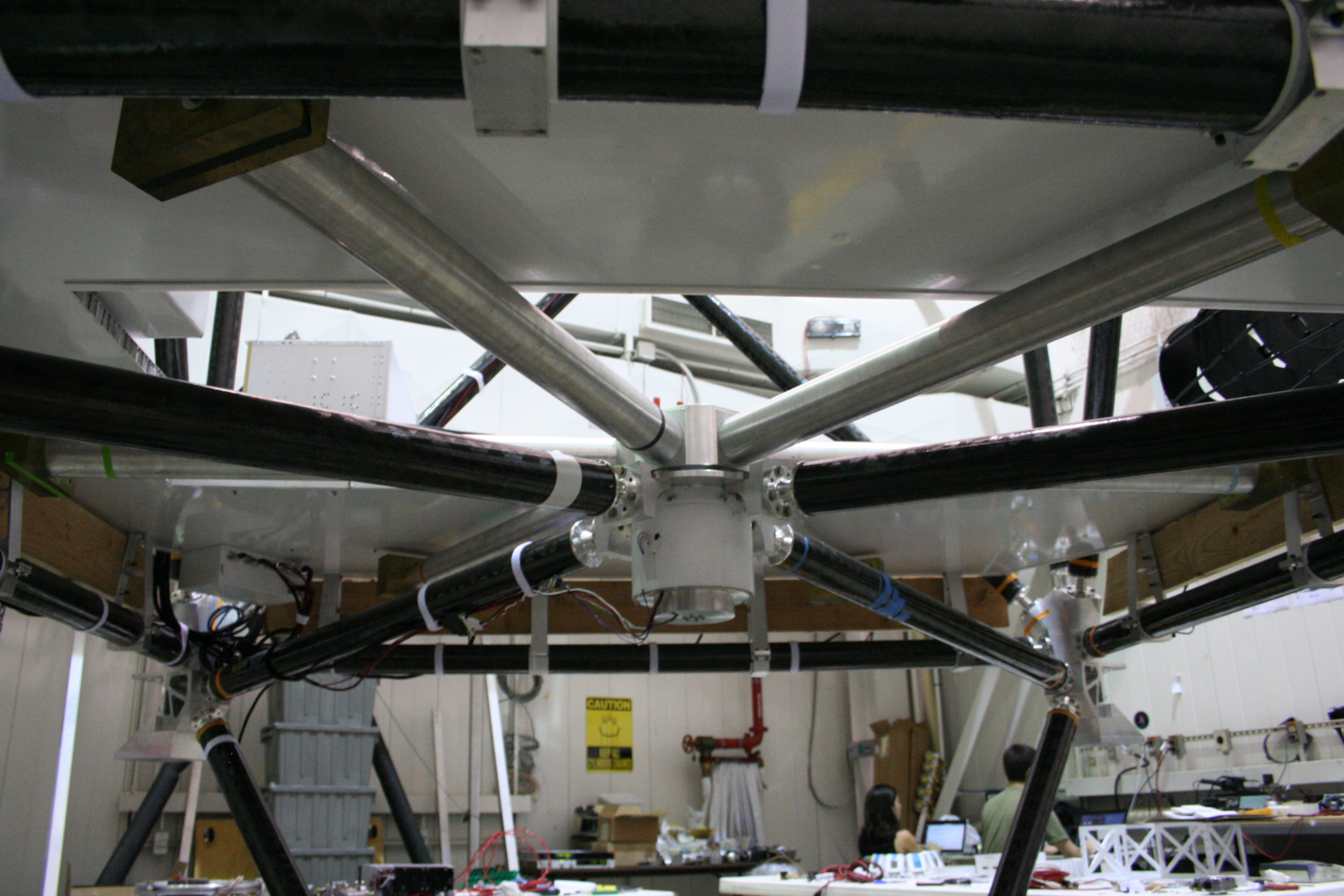}
\caption{\label{fig:wheel} \emph{Left} -- A SolidWorks rendering of the \Spider~reaction wheel, showing the bricks, spokes, hub, and motor. The spokes have been made transparent in order to reveal the threaded rods. \emph{Right} -- a photograph of the reaction wheel as it is mounted on the underside of the gondola.}

\end{figure}

The reaction wheel consists of six 7 kg brass bricks connected to a central hub by 1 m long spokes (Figure \ref{fig:wheel}). These spokes are 6061-T6 aluminum pipes with a $4''$ outer diameter (OD) and a $0.125''$ wall thickness. Structural support is provided by 3/8-16 threaded rods running through the bricks and spokes, and terminating inside the central hub. Mass is concentrated around the outside of the wheel in order to maximize its rotational inertia at a given weight. The reaction wheel's moment of inertia around its spin axis is 44.77 kg$\cdot$m$^2$, and its mass is 50.34 kg, including the motor, which extends below the wheel.

\subsubsection{The Motor Torque Control Loop}
\label{sec:rwctrl}

The reaction wheel is driven by a K178200-6Y1 brushless DC motor from Parker Bayside Motion (Table~\ref{tab:motors}). This frameless motor has been installed in a custom-designed motor housing that includes a resolver for feedback sensing. The resolver is an analog rotary position sensor that determines the angular position and speed of the rotor shaft relative to the stator windings. This sensor is necessary for correct commutation of the motor. Motor current is driven by a DPRALTR-060B080 digital servo amplifier from Advanced Motion Controls (AMC). This servo drive carries out digital commutation based on the resolver feedback. It can drive  current through the motor windings up to a maximum of 60 A. The servo drive regulates the output current, executing the lowest-level control loop in the reaction wheel control system.

\begin{figure}[htbp!]
\begin{center}
\includegraphics[width=\textwidth]{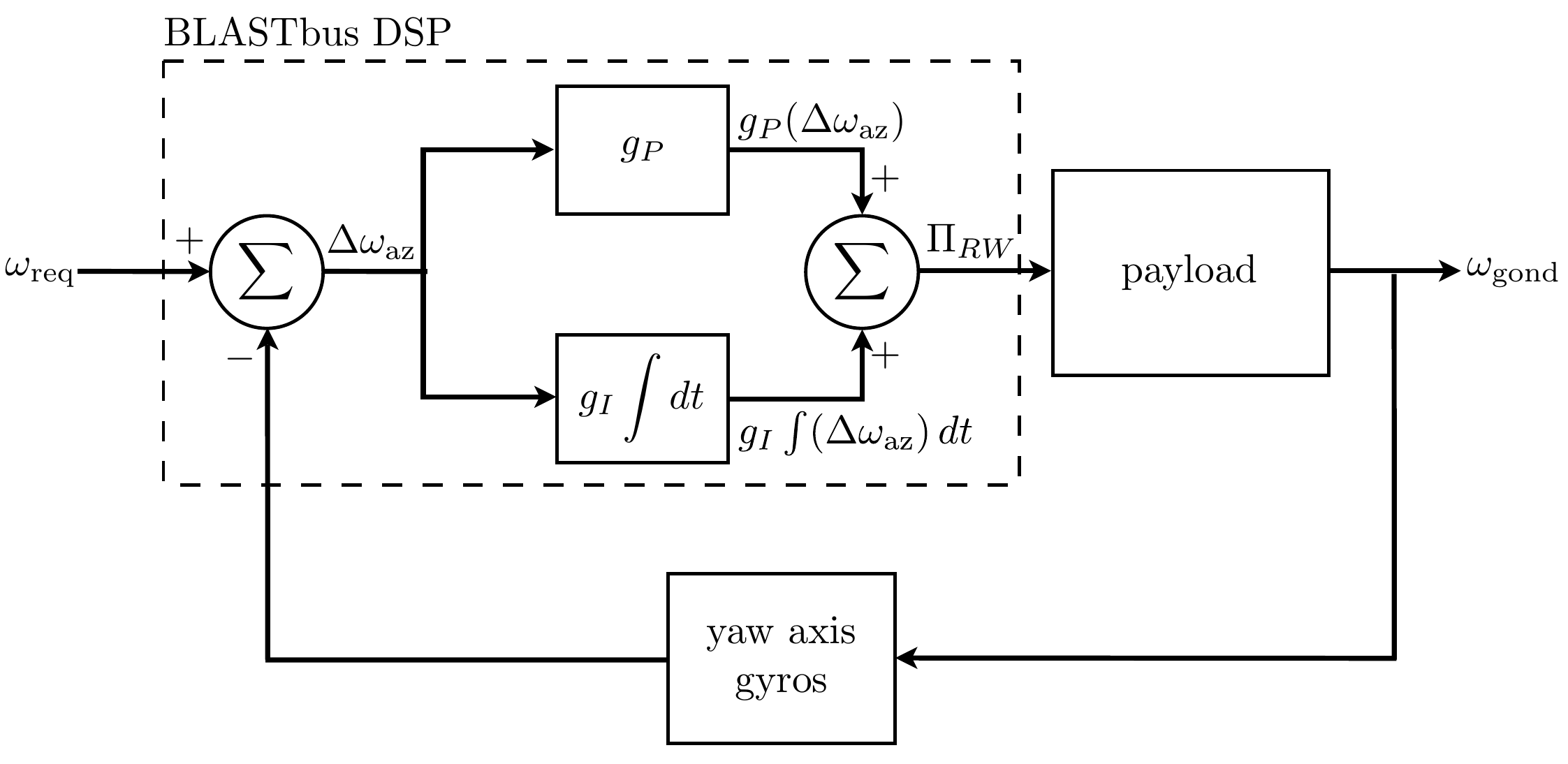}
\caption{\label{fig:rwloop} A block diagram of the reaction wheel (RW) motor torque control loop}
\end{center}
\end{figure}

\begin{figure}[htbp!]
\begin{center}
\includegraphics[width=\textwidth]{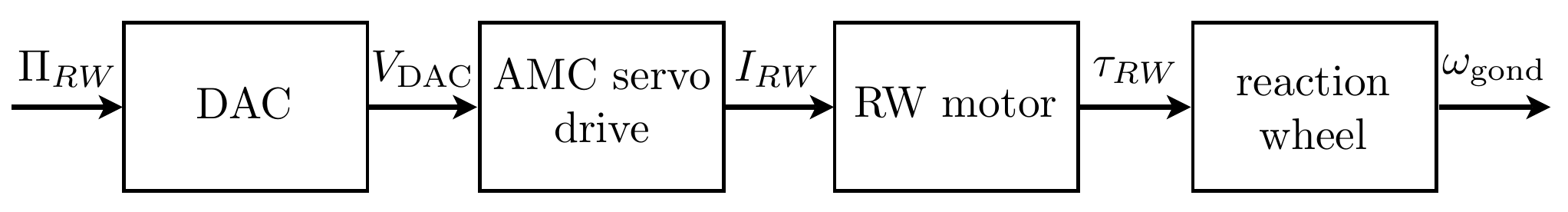}
\caption{\label{fig:block} Expanded diagram of the block labelled ``payload'' in Figure~\ref{fig:rwloop}, illustrating how the DSP control loop ultimately controls reaction wheel torque by setting the DAC level $\Pi_{RW}$ based on the az velocity error.}
\end{center}
\end{figure}

A voltage is supplied to one of the servo drive's analog inputs by a $\pm$5~V Digital-to-Analog Converter (DAC). The drive interprets this as an output current request using a proportionality constant of 3.6 A/V. This determines the motor torque\footnote{For these motors, $\tau = K_t I$, where $K_t$ is the torque constant (Table~\ref{tab:motors})} (Figure~\ref{fig:block}). The input DAC level is determined by a Proportional, Integral (PI) negative feedback control loop (Figure~\ref{fig:rwloop}).

As shown in Figure~\ref{fig:rwloop}, the feedback sensors for this control loop are \Spider's yaw axis gyroscopes. \Spider~has six KVH DSP-3000 digital fibre optic rate gyroscopes: two for each of the yaw, pitch, and roll rotation axes. The gyros measure the payload angular velocity in each axis, relative to an inertial frame. Ignoring pendulations, rotation around the yaw axis corresponds to motion of the payload in azimuth. In the control system, the yaw axis gyros are sampled at approximately 13~kHz by a digital signal processor (DSP) on one of the BLASTbus motherboards in the ACS. At this rate, a real-time process on the DSP subtracts this measured angular velocity, $\omega_\mathrm{gond}$, from the requested azimuthal angular velocity $\omega_\mathrm{req}$, to produce the az velocity error:
\begin{equation}
\Delta \omega_\mathrm{az} \equiv \omega_\mathrm{req} - \omega_\mathrm{gond}.
\end{equation}
The DSP then computes the two control terms. The $P$ term is proportional to $\Delta \omega_\mathrm{az}$ , so that the reaction wheel motor torque increases in proportion to the velocity error that it is trying to correct. The $I$ term is proportional to the integral of $\Delta \omega_\mathrm{az}$, which helps reduce steady-state error:
\begin{equation}
\Pi_{RW} = g_P (\Delta \omega_\mathrm{az}) + g_I \int (\Delta \omega_\mathrm{az})\,dt.
\end{equation}
The result, $\Pi_{RW}$ is the 16-bit DAC digital input level. Therefore, based on the velocity error, the DSP control loop ultimately determines the reaction wheel motor torque $\tau_{RW}$ (Figure~\ref{fig:block}). The values of the reaction wheel gains $g_P$ and $g_I$ can be commanded in flight, and will be tuned shortly after the payload reaches float altitude ($\gtrsim$ 120~000 ft), during the period when line-of-sight commanding is possible.  

With this control loop in operation on the DSP, the payload is a velocity-commandable system, from the point of view of the flight computers. On the flight computers, \Spider's master control program, known as pcm, computes $\omega_\mathrm{req}$ at approximately 120~Hz, and its value is communicated to the ACS over the BLASTbus. More complex pointing and scanning motions are achieved by programming pcm to vary $\omega_\mathrm{req}$ with time (see Section~\ref{sec:obs}). 
 
\subsection{The Pivot}
\label{sec:piv}

\subsubsection{Mechanical Design}
\label{sec:pivdesign}

\begin{figure}[htbp!]
\begin{center}
\includegraphics[width=0.5\textwidth]{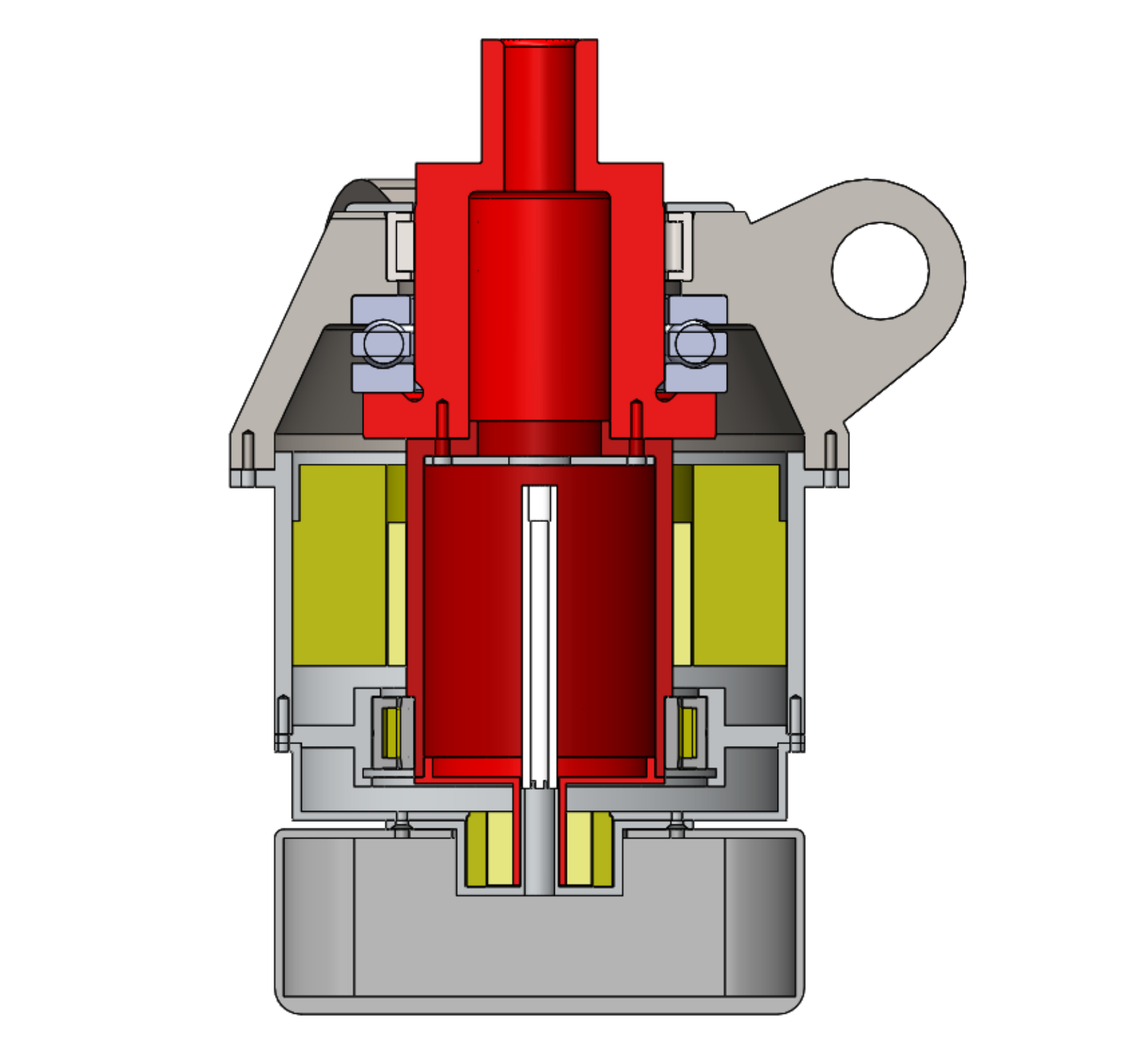}
\includegraphics[width=0.4\textwidth]{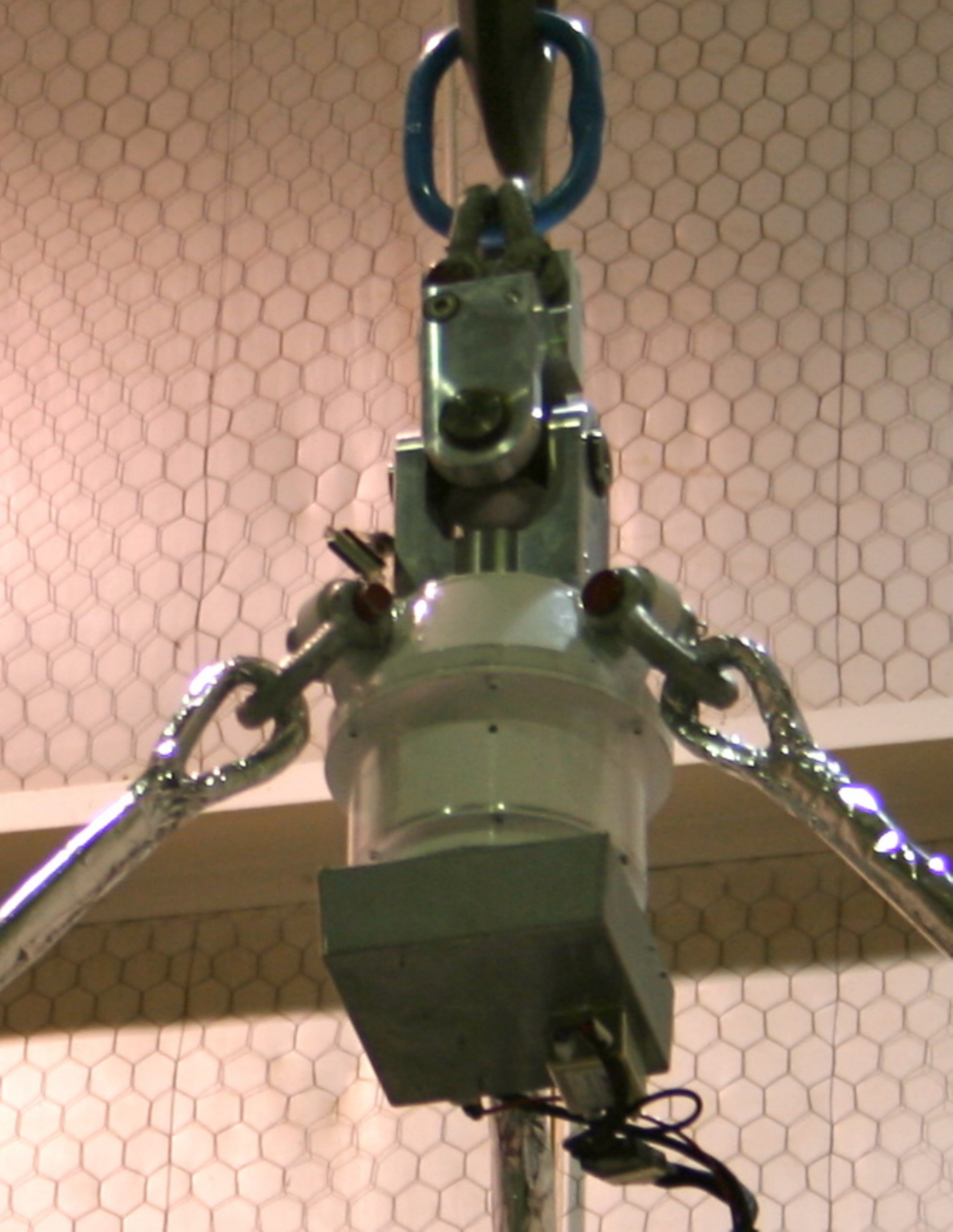}
\caption{\label{fig:pivmech} \textit{Left} -- A cross-sectional view of a SolidWorks model of the pivot motor. \textit{Right} -- A photograph of the pivot showing, from top to bottom, the universal joint that connects to the flight train, the motor casing (white), and the pivot control box. The three cables from which the gondola is suspended are also visible. }
\end{center}
\end{figure}

As shown in the left panel of Figure~\ref{fig:pivmech}, the pivot design incorporates three bearings. At the top of the pivot is a steel section of the casing that has welded tabs to which the ropes that suspend the gondola are attached. The pivot rotor shaft (red) is supported within this section of the casing by an SKF 51218 thrust ball bearing. This bearing supports the entire weight of the payload. Just above this bearing is an SKF NK 90/25 needle roller bearing. Another identical needle roller bearing supports the rotor farther down, in the cylindrical section of the casing where the motor windings are located. The primary purpose of the needle bearings is alignment. The resolver is visible as a small cylindrical protrusion at the very bottom of the casing, extending partially into the control box.

\begin{table}[htbp!]
\caption{Properties of the reaction wheel and pivot brushless DC motors from Parker Bayside Motion. The physical parameters are the same for both models: only the winding constants differ. To ensure proper operation, the correct winding constants must be stored in the firmware of the each motor's servo drive.} 
\label{tab:motors}
\begin{center}       
\begin{tabular}{lcc}
\hline
\rule[-1ex]{0pt}{3.5ex} & \textbf{Reaction Wheel} & \textbf{Pivot}  \\
\hline 
\hline 
\rule[-1ex]{0pt}{3.5ex}  Model No. & K178200-6Y1 & K178200-8Y1 \\~\\
\rule[-1ex]{0pt}{3.5ex}  \textit{Physical Parameters} &\\
\hline
\rule[-1ex]{0pt}{3.5ex}  Maximum Mechanical speed [rpm] & \multicolumn{2}{c}{6000}\\ 
\rule[-1ex]{0pt}{3.5ex}  Stall Torque Continuous [N$\cdot$m] & \multicolumn{2}{c}{25.74}\\ 
\rule[-1ex]{0pt}{3.5ex}  Maximum Winding Temperature [$^\circ$C]  & \multicolumn{2}{c}{155}\\
\rule[-1ex]{0pt}{3.5ex}  Rotor Inertia [kg$\cdot$m$^2$]  & \multicolumn{2}{c}{1.8 $\times$ 10$^{-3}$}  \\
\rule[-1ex]{0pt}{3.5ex}  Number of Rotor Magnet Poles & \multicolumn{2}{c}{18}  \\
\rule[-1ex]{0pt}{3.5ex}  Mass [kg] &  \multicolumn{2}{c}{6.34} \\~\\
\rule[-1ex]{0pt}{3.5ex}  \textit{Winding Constants} & &\\
\hline
\rule[-1ex]{0pt}{3.5ex}  Stall Current Continuous [A$_\mathrm{rms}$] & 12.9 & 8.15  \\
\rule[-1ex]{0pt}{3.5ex}  Peak Current [A$_\mathrm{rms}$] & 40.9 & 25.76  \\
\rule[-1ex]{0pt}{3.5ex}  Voltage Constant $K_b$ [V/(rad$\cdot$s$^{-1}$)] & 1.639 & 2.595  \\
\rule[-1ex]{0pt}{3.5ex}  Torque Constant $K_t$ [N$\cdot$m/A$_\mathrm{rms}$] & 2.007 & 3.178  \\
\rule[-1ex]{0pt}{3.5ex}  Resistance [$\Omega$] & 0.6857 & 1.7  \\
\rule[-1ex]{0pt}{3.5ex}  Inductance [mH] & 6.118 & 15.3  \\
\hline
\end{tabular}
\end{center}
\end{table}

\subsubsection{The Motor Velocity Control Loop}
\label{sec:piv_vel}

The pivot is driven by a K178200-8Y1 brushless DC motor from Parker Bayside Motion. This motor has the same physical dimensions as the reaction wheel motor (178~mm OD), but with a different set of windings (Table~\ref{tab:motors}). The system driving the pivot motor is similar to the reaction wheel system described in Section~\ref{sec:rwctrl}, with a few important differences. First, the pivot's DPRALTR-020B080 AMC servo drive has a peak current rating of only 20 A, vs.~60 A for the reaction wheel's. Second, the pivot control loop is implemented on the flight computers as part of pcm, instead of on a DSP in the ACS. Therefore, it operates at a lower rate ($\sim$120~Hz). Third, the pivot's AMC servo drive has been programmed to interpret the pivot DAC voltage as a \emph{velocity request}, rather than a current request. The analog input scaling is approximately 33~rpm/V. The drive attempts to servo the pivot rotor velocity to the requested value, based on resolver feedback. Operating the pivot in velocity mode rather than torque mode is a significant departure from the pivot control in previous balloon payloads. In torque mode, abrupt pivot motions were found to drive pitch and roll pendulation modes of the payload, especially during scans as wide and fast as \Spider's (Section~\ref{sec:obs}). Since static friction is considerably larger than rolling friction in the pivot, motor current would build up until the static friction was overcome, resulting in rapid motion. This problem is avoided with velocity control, which ensures smooth and continuous motion of the pivot.

In BLASTPol, whose pivot operated in torque mode, the control terms included the following
\begin{equation}
\label{eq:oldpiv}
\tau_\mathrm{piv} = g_1 \Delta \omega_{RW} + g_3 \Delta \omega_\mathrm{az},
\end{equation}
where $\Delta \omega_{RW} \equiv \omega_{SP} - \omega_{RW}$. The first pivot control term attempts to servo the reaction wheel rotation speed to a setpoint value, $\omega_{SP}$, by providing a torque proportional to the error between this setpoint and the measured reaction wheel speed $\omega_{RW}$\footnote{The measured reaction wheel speed comes from the resolver, and is read by the flight computers from the reaction wheel motor's AMC servo drive over RS-232.}. The second control term helps the payload to scan by providing a torque proportional to the az velocity error. On \Spider, the second term of Eq.~\ref{eq:oldpiv} was found to cause random pivot motions due to amplification of the gyro noise present in $\Delta \omega_\mathrm{az}$. This term was replaced with $g_3\alpha_\mathrm{req}$, where $\alpha_\mathrm{req}$ is the requested or ``theoretical'' payload angular acceleration, in azimuth. During a scan, $\alpha_\mathrm{req}$ varies deterministically with azimuth.

When these measures failed to prevent sudden pivot motions, velocity control was implemented. It was necessary to translate the terms of Eq.~\ref{eq:oldpiv} into equivalent expressions for $\omega_\mathrm{piv}$, the pivot rotation rate. The pivot provides torque by twisting the flight train, which can be considered a torsional spring. Therefore, $\tau_\mathrm{piv}  = k\theta_\mathrm{piv}$, where $k$ is the effective flight train torsional spring constant\footnote{The effective spring constant of the flight train used by CSBF has been determined theoretically to have a value of 0.4~N$\cdot$m/deg, and this agrees with measurements done during the 2012 flight of BLASTPol~\cite{lmf2013}.}, and $\theta_\mathrm{piv}$ is the rotation angle of the pivot, measured relative to the angle of zero twist in the flight train. Differentiating both sides of the equation yields $\dot{\tau}_\mathrm{piv} = k\omega_\mathrm{piv}$. Substituting in the expression $g_3\alpha_\mathrm{req}$ for $\tau_\mathrm{piv}$, we obtain $\omega_\mathrm{piv} = (g_3/k)\dot{\alpha}_\mathrm{req}$ as an equivalent of the second control term in velocity mode. During a scan, \Spider's az varies sinusoidally with time. Therefore, so do all of its derivatives. In particular, for a sinusoidal scan profile, $\dot{\alpha}_\mathrm{req} \propto -\omega_\mathrm{req}$. The scan control term can therefore be implemented by setting $\omega_\mathrm{piv} \propto -\omega_\mathrm{req}$ with some proportional gain.

Two additional control terms were implemented in velocity mode to servo the reaction wheel to the setpoint speed. Note that in \Spider, $\omega_{SP}$ is typically set to zero. The width and angular acceleration of the scan necessitate a large swing in reaction wheel speed, centred on 0 deg/s. Therefore $\Delta \omega_{RW} = \omega_{RW}$, and the two control terms are given by
\begin{equation}
\label{eq:newpiv}
\omega_\mathrm{piv} = g_1 \omega_{RW} + g_2 \tau_{RW}.
\end{equation}
The effect of these terms can be understood by considering the simple case in which the payload in is \emph{stop mode}, meaning that the pointing system is trying to servo to zero speed in azimuth. Under these conditions, there is no net torque on the payload from the pointing motors: $\tau_{RW}+\tau_\mathrm{piv} = 0$. Differentiating both sides, we obtain $\dot{\tau}_\mathrm{piv} = k\omega_\mathrm{piv} = -I\ddot{\omega}_{RW}$ where $I$ is the reaction wheel moment of inertia. Substituting in the expression for $\omega_\mathrm{piv}$ from Eq.~\ref{eq:newpiv} and re-arranging, we obtain
\begin{equation}
\label{eq:shm}
\ddot{\omega}_{RW} + kg_2\dot{\omega}_{RW} + \frac{k}{I}g_1\omega_{RW} = 0.
\end{equation}
The dynamical equation for the reaction wheel angular velocity is that of a damped harmonic oscillator. The strength of the damping depends on $k$ and the gain $g_2$. The undamped frequency of oscillation is higher for stiffer $k$ and lower for larger $I$. A non-zero net torque on the payload acts as a driving term. Figure~\ref{fig:shm} below shows a measurement of the reaction wheel speed taken during lab testing that exhibits this behaviour.

\begin{figure}[hbtp!]
\begin{center}
\includegraphics[width=\textwidth]{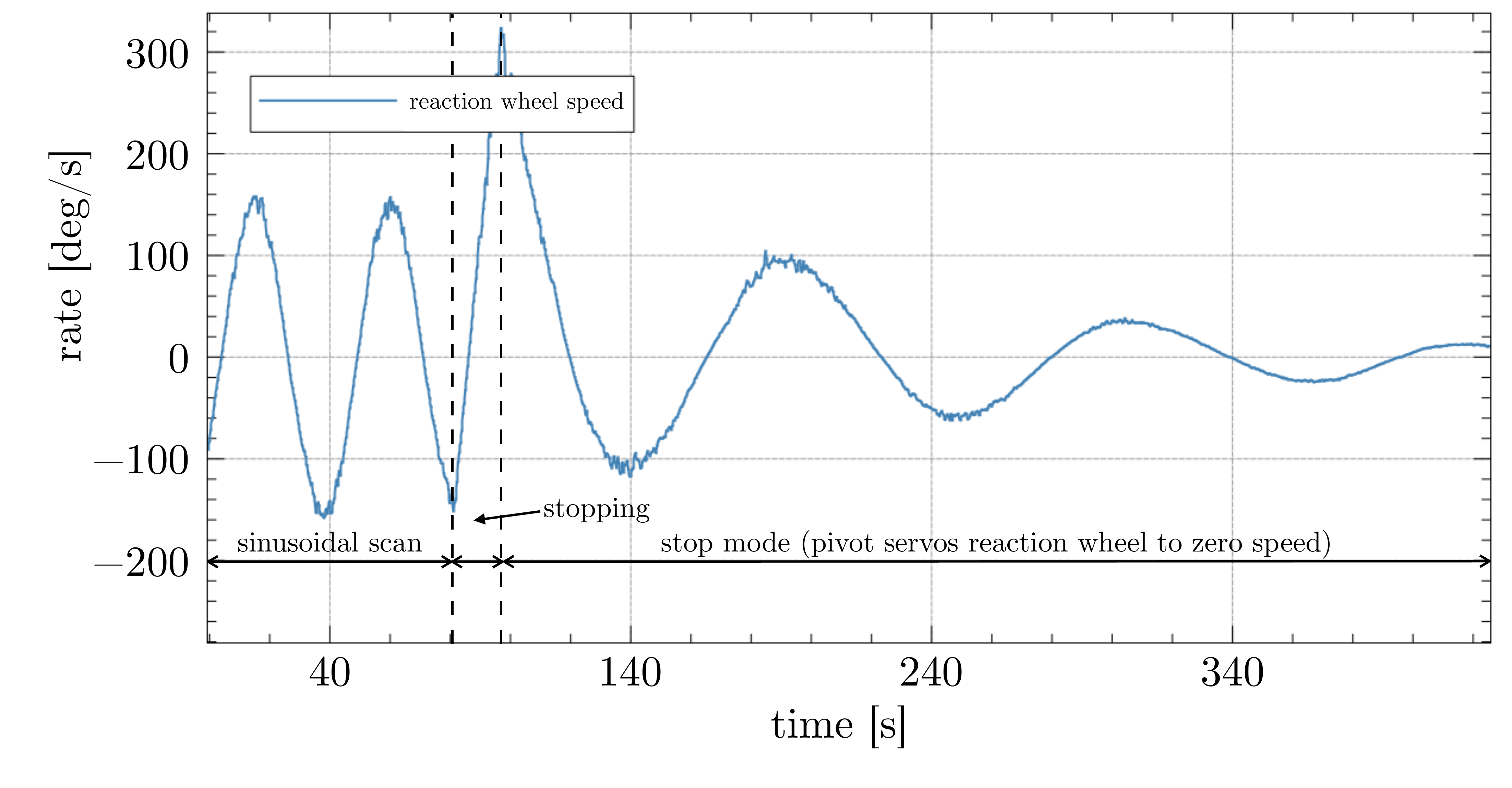}
\caption{\label{fig:shm} Reaction wheel speed vs.~time during scan tests conducted with the payload suspended from a simulated flight train. In this plot, the gondola has just gone into stop mode from a sinusoidal scan, one period of which is visible on the left. The large angular acceleration involved perturbs the system of Eq.~\ref{eq:shm}, causing an increase in reaction wheel speed that then damps out, returning to the setpoint value.}
\end{center}
\end{figure}

Combining the two reaction wheel control terms (Eq.~\ref{eq:newpiv}) with the scanning term results in an overall pivot control loop with three proportional gain terms: $\omega_\mathrm{piv} = g_1 \omega_{RW} + g_2 \tau_{RW} - g_3 \omega_\mathrm{req}$. This equation is in terms of the dynamical quantities of interest. However, as with the reaction wheel control loop on the DSP, what the pivot control loop in pcm actually computes is $\Pi_\mathrm{piv}$, the 16-bit pivot DAC level. This is translated into $\omega_\mathrm{piv}$ by the combination of the pivot DAC and the AMC servo drive. Furthermore,  $\tau_{RW}$ is not measured directly, but estimated using the \emph{reaction wheel's} DAC level $\Pi_{RW}$, which can be regarded as the \emph{commanded} reaction wheel torque. Therefore, what pcm computes is more properly expressed as
\begin{equation}
\Pi_\mathrm{piv} = g_1 \omega_{RW} + g_2 \Pi_{RW} - g_3 \omega_\mathrm{req}.
\end{equation}
The parameters $g_1$, $g_2$, $g_3$, and $\omega_{SP}$ can be commanded in flight. The pivot gains will be tuned shortly after the payload reaches float altitude. Pivot velocity control showed excellent performance during scan tests conducted with a simulated flight train that has realistic dynamical properties. However, it is still true that this control system has never successfully flown before. As a contingency against the inability to tune the system at float, we have implemented a command that reverts the pivot's AMC servo drive to torque mode, and switches the pivot control terms in pcm over to those in Eq.~\ref{eq:oldpiv}.

\section{THE ELEVATION DRIVE}
\label{sec:el}

\subsection{Mechanical Design}

\begin{figure}[hbtp!]
\begin{center}
\includegraphics[width=0.75\textwidth]{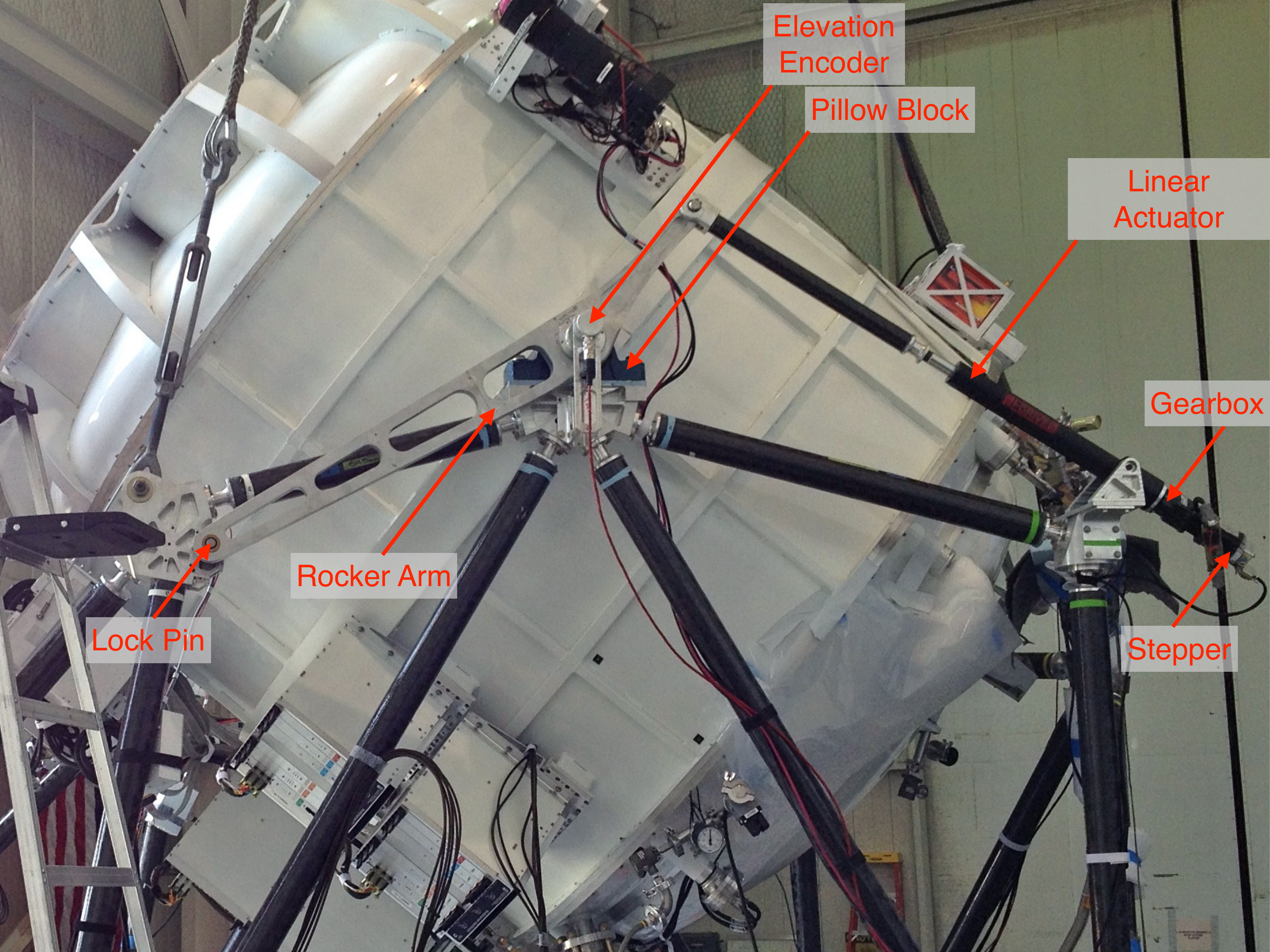}
\caption{\label{fig:eldrive} A photograph of the port side of the cryostat and gondola, with elements of the elevation drive labelled. These elevation drive components are repeated on the starboard side.}
\end{center}
\end{figure}

The cryostat mounts to the gondola frame as shown in Figure~\ref{fig:eldrive}. Trunnions mounted to the sides of the cryostat engage with SKF FSYE-3-NH pillow block bearings that rest atop the gondola joints at the centre of the port and starboard sides of the frame. The pillow blocks provide an axis for the cryostat to rotate in elevation (el). Aluminum rocker arms bolt to the outside ends of the trunnions, such that they are rigidly attached to the cryostat. Linear actuators mounted farther back on the gondola frame push on the rocker arms. Extension of the linear actuators lowers the cryostat's elevation, while retraction raises it. This system provides the mechanical advantage necessary to rotate the 3500~lb cryostat, even if it is unbalanced. The centre of mass of the cryostat will shift as cryogens boil off.

The linear actuators are VecTac E-Drive VT209-12 ball screw actuators rated for 900 lb$_\mathrm{f}$ of thrust. This model has a 12$''$ throw. The input shafts of these actuators are each connected with a flex coupling to a Stober P221SPR0070MT ServoFit\texttrademark~Precision Planetary Gearhead with a 7:1 gear ratio. This gearbox couples to a NEMA 23 inline brake that is spring-actuated to prevent motion when the system is not powered by 24 VDC. The brake couples to a Cool Muscle CM1-C-23L20C stepper motor from Myostat Motion Control Inc. The couplings between the gearbox and the inline brake, and between the brake and the stepper motor, are secured with shaft collars. 

Elevation feedback sensing is provided by two Encoder Technology EA58-S absolute electro-optical encoders. These are rigidly mounted to the gondola frame, and their output shafts connect with flex couplings to shafts protruding from the sides of each rocker arm, directly inline with the elevation axis. Thus, the encoders measure the cryostat elevation angle relative to the gondola. The encoders have a 16-bit resolution, resulting in a minimum el step measurement of 0.0055$^\circ$.

\subsection{Motor Drive and Power}

Each Cool Muscle stepper unit includes a built-in motor controller and a magnetic encoder for high resolution feedback on the rotor position, making each unit a fully-integrated servo system. By operating closed-loop with fast feedback, the motors avoid the drawbacks of other steppers, such as vibration and missed steps. Steppers can also be reliably commanded to any velocity simply by controlling the step rate. This feature enables a position control loop in which velocity is commanded based on the position error (Section~\ref{sec:elctrl}). This type of control is suitable given that the only feedback sensing in elevation is the position angle from the encoders. An alternative would have been to use brushless DC torque motors controlled with pulse-width modulation (PWM). However, in the absence of fast velocity feedback, it is difficult to tune such a system to servo to position reliably, especially given the static friction in the system and the unbalanced load.

The Cool Muscles' controllers are commanded in pulse-per-step mode with two opto-isolated step and direction inputs. The step input receives a square wave pulse train whose frequency determines the step rate. The pulse frequency is computed by pcm as described in Section~\ref{sec:elctrl}. This frequency is communicated to the ACS over the BLASTbus at $\sim$120~Hz. The square wave output is generated by an Altera FPGA on one of the BLASTbus motherboards in the ACS. The system imposes a maximum pulse frequency of 10~kHz. 

The power supply to the elevation drive required some careful consideration. Most 24~V components on the experiment accept an input range of 18 to 36~VDC, and can therefore be powered directly from the flight batteries. The Cool Muscle steppers are an exception, requiring an input of 24~V $\pm$ 10\%. As a result, each motor is powered by a Vicor DC-to-DC converter (DC-DC). Since Vicors cannot handle reverse current, which is produced by inductive loads such as the motors, two 35~$\Omega$ Dale 50 W power resistors in series (70~$\Omega$ total) are placed across the output of each DC-DC, in parallel with the motor, biasing the output current to be positive. 

The output 24~VDC from each Vicor is provided in parallel to both a Cool Muscle stepper and its inline brake. Powering the brakes causes them to disengage, enabling motion. The Cool Muscle controllers take time to power on and energize the motor windings, providing holding torque. Therefore, a time delay was introduced between powering the motors and disengaging the brakes, in order to prevent the cryostat from falling. The two components remain powered by the same 24~VDC circuit, and the delay is implemented in hardware, so that the brake is never disengaged when the motor is unpowered. At the motor, the connection to the inline brake is interrupted by a Crydom CMX60D20 solid state relay whose input is controlled by a simple RC circuit. When 24~VDC is applied, it reaches the motors immediately, and the inline brakes approximately 300~ms later.

\subsection{Position Control Algorithm}
\label{sec:elctrl}

In every scan mode, pcm computes a requested elevation angle $\theta_\mathrm{req}$. At the BLASTbus data rate of $\sim$120~Hz, the elevation position control routine in pcm computes a rotation rate $\omega_\mathrm{el}$ for the elevation drive as follows
\begin{equation}
\omega_\mathrm{el} = \mathrm{sgn}(\Delta \theta)g_\mathrm{el}\sqrt{|\Delta \theta|},
\end{equation}
where the position error $\Delta \theta \equiv \theta_\mathrm{req} - \theta_\mathrm{enc}$, with $\theta_\mathrm{enc}$ being the mean of the starboard and port elevation encoder readings. The square root velocity-position profile corresponds to a constant negative acceleration to zero speed as the measured elevation approaches the target value. The magnitude of the acceleration depends on the gain $g_\mathrm{el}$, whose value can be commanded in flight, but has already been tuned during laboratory tests. When $|\Delta \theta| \leq 0.02^\circ$, pcm simply sets $\omega_\mathrm{el}$ to zero.

The computed rotation rate is limited to $|\omega_\mathrm{el}| \leq \omega_\mathrm{max}$, where $\omega_\mathrm{max}$ is the rate of change of elevation corresponding to the maximum motor pulse frequency of 10~kHz. The gain is set high enough that the pulse rate saturates to this value during large elevation slews. To convert between motor pulse frequency and rotation rate in el, pcm must first compute
\begin{equation}
\label{eq:thetaell}
\omega_\mathrm{el} = \frac{d\theta}{dt} = \frac{d\theta}{d\ell}\frac{d\ell}{dt},
\end{equation}
where $\ell$ is the linear actuator extension. Due to the geometry of the system (Figure~\ref{fig:eldrive}), $\frac{d\theta}{d\ell}$ varies with $\theta$ and pcm computes it at $\sim$120~Hz using the measured elevation $\theta_\mathrm{enc}$. Having computed the rate of change of linear actuator extension, pcm can determine the pulse frequency as follows:
\begin{equation}
\label{eq:ellpulse}
\frac{d\ell}{dt} =  (\mathrm{gear~ratio}) \cdot \left(\frac{\mathrm{linear~actuator~thrust~(mm)}}{\mathrm{rotation}}\right) \cdot \left(\frac{\mathrm{stepper~rotations}}{\mathrm{pulse}}\right) \cdot f_\mathrm{el},
\end{equation} 
where $f_\mathrm{el}$ is the pulse frequency. The gear ratio is 7:1, the actuator undergoes 5 rotations per inch of extension, and the number of motor steps per full rotation is set to 5000 in the firmware of the Cool Muscle controllers. Given these numbers, the maximum pulse frequency of 10~kHz results in motion of the linear actuators at 1.45~mm/s. Given the time it would take to cover the full 12$''$ range of the linear actuator at that rate, the \emph{average} elevation speed over the full mechanical el range of 11.6$^\circ$ to 56.4$^\circ$ is $\langle\omega_\mathrm{el}\rangle \approx 0.2$~deg/s at the maximum pulse rate. 

\section{OBSERVING MODES}
\label{sec:obs}

As discussed in Section~\ref{sec:rwctrl}, the fast reaction wheel motor torque control loop on the DSP servos the payload's azimuthal angular velocity to a requested value $\omega_\mathrm{req}$. Higher-level control algorithms, implemented in pcm on the flight computers, produce various scanning and pointing motions by varying $\omega_\mathrm{req}$ as a function of time or other observables. The most important scan modes are discussed below, including the main strategy planned for carrying out CMB observations during flight.

\subsection{Sinusoidal Azimuth Scan}
\label{sec:sinescan}
A scan mode in pcm that carries out a sinusoidal scan in azimuth at a constant elevation was developed for laboratory tests of the pointing system, and as a subcomponent of the science scan mode (Section~\ref{sec:spiderobs}). A sinusoidal profile was chosen in order to make it possible, within the limitations of motor torque, to carry out scans of the required speed and width. In particular, \Spider~must scan fast enough to modulate the sky signal into frequencies above the $1/f$ knee of the detector noise spectra. A scan with a constant speed equal to the peak speed of the sinusoidal scan would require more acceleration at scan turn-arounds than the pointing motors can provide. The sinusoidal scan is completely defined by three parameters: the scan centre, $\phi_c$, the scan amplitude $A$, and the peak az angular acceleration, $\alpha_\mathrm{pk}$. During a scan, the variation of az angle $\phi$ with time is of the form
\begin{equation}
\label{eq:phi}
\phi(t) = \phi_c  - A\cos\left(\frac{2\pi}{T}t\right),
\end{equation}
where $T$ is the scan period, and the phase of the scan has been chosen arbitrarily. The corresponding expressions for angular velocity and acceleration are
\begin{equation}
\label{eq:omega}
\dot{\phi}(t) = \omega(t) = \frac{2\pi}{T}A\sin\left(\frac{2\pi}{T}t\right)
\end{equation}
and
\begin{equation}
\label{eq:alpha}
\ddot{\phi}(t) = \alpha(t) = \frac{4\pi^2}{T^2}A\cos\left(\frac{2\pi}{T}t\right).
\end{equation}
The angular velocity and acceleration can also be written as $\omega(t) = \omega_\mathrm{pk} \sin\left(\frac{2\pi}{T}t\right)$ and $\alpha(t) = \alpha_\mathrm{pk} \cos\left(\frac{2\pi}{T}t\right)$. Comparing these two equations with Eqs.~\ref{eq:omega} and~\ref{eq:alpha} above, it can be shown that
\begin{equation}
T = 2\pi\sqrt{\frac{A}{\alpha_\mathrm{pk}}}
\end{equation}
and
\begin{equation}
\omega_\mathrm{pk} = \sqrt{\alpha_\mathrm{pk} A}.
\end{equation}

Since the available pivot and reaction wheel motor torque ultimately limits the maximum payload angular acceleration, $A$ and $\alpha_\mathrm{pk}$ are the base parameters of the scan mode. The values of $\omega_\mathrm{pk}$ and $T$ result from the values of these base parameters, rather than being specified directly. This reduces the probability of a sinusoidal scan with an unattainable value of $\alpha_\mathrm{pk}$ being commanded. For the purposes of testing, $\alpha_\mathrm{pk} = 0.8~\mathrm{deg} \cdot \mathrm{s}^{-2}$ and $A = 45^\circ$ have long been considered the canonical ``flight-like'' scan parameter values, resulting in $T = 47$~s and $\omega_\mathrm{pk} = 6~\mathrm{deg} \cdot \mathrm{s}^{-1}$. However, during flight, in the science scan mode, $A$ varies with time based on other parameters (Section~\ref{sec:spiderobs}).  

\begin{figure}[hbtp!]
\begin{center}
\includegraphics[width=\textwidth]{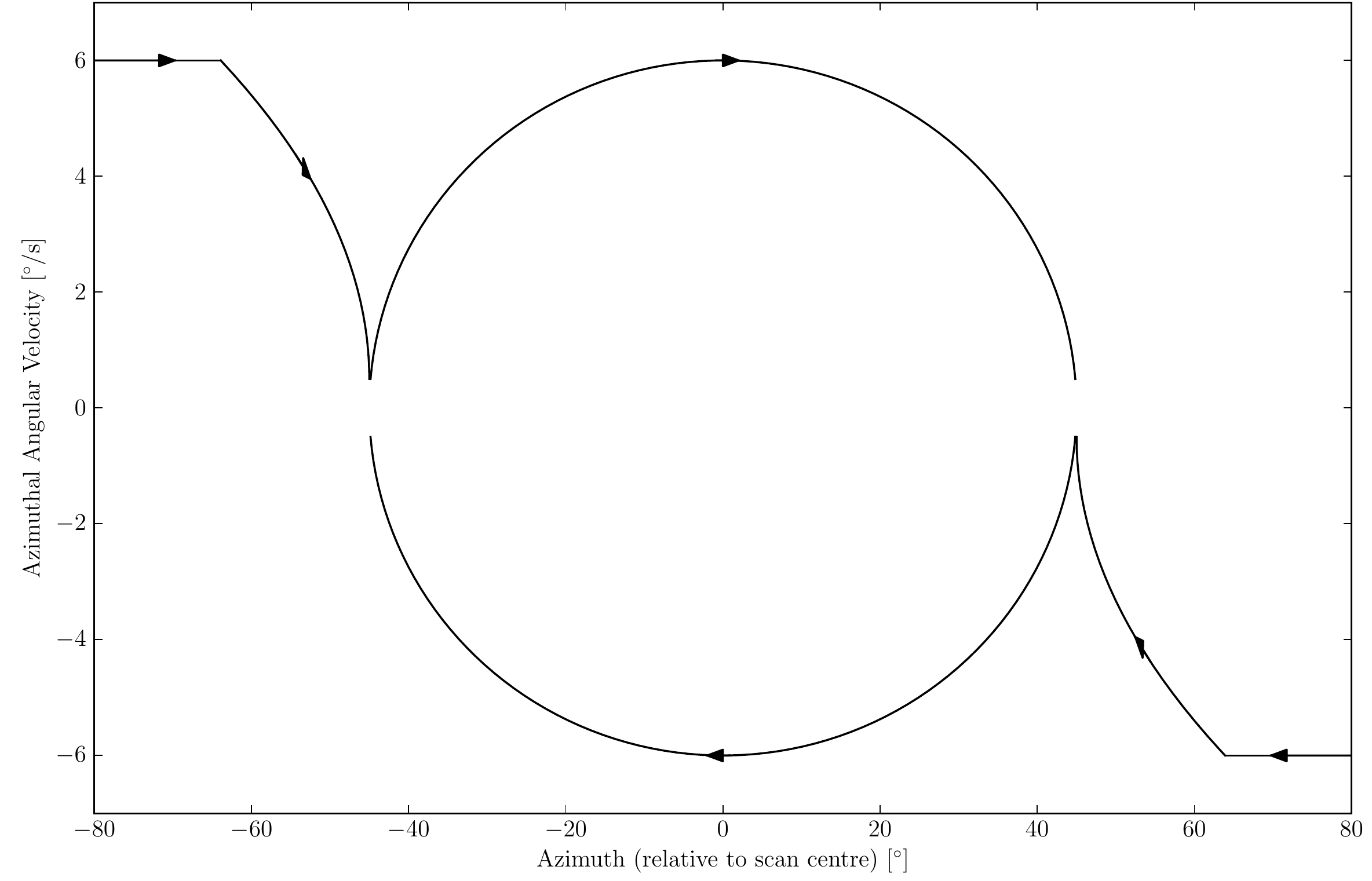}
\caption{\label{fig:ellipse} An example of $\omega_\mathrm{req}$ vs.~az (relative to the scan centre) as determined by the sinusoidal scan algorithm in pcm. The canonical values of $A=45^\circ$ and $\omega_\mathrm{pk} = 6~\mathrm{deg/s}$ were used. When the payload azimuth is beyond the scan endpoints on either side, the velocity follows a square root profile with acceleration $\alpha_\mathrm{pk}$, up to a maximum speed of $\omega_\mathrm{pk}$.  Within the scan endpoints, the velocity follows the ellipse corresponding to a sinusoidal scan (Eq.~\ref{eq:ellipse}). The algorithm follows the elliptical profile until $|\omega_\mathrm{req}| < \omega_\mathrm{min}$, which occurs at a distance of $\Delta \phi_\mathrm{turn}$ before the turn-around. Within this distance of a scan endpoint, the speed is constant at $\omega_\mathrm{min}$; $\omega_\mathrm{req}$ simply flips sign from $\pm\omega_\mathrm{min}$ to $\mp\omega_\mathrm{min}$. Therefore, the algorithm jumps from the top to the bottom elliptical branch, or vice versa. In pcm, $\omega_\mathrm{min}$ = 0.05~deg/s, which is considered to be the smallest reliably-measurable speed given gyro noise and offsets. In the figure,  $\omega_\mathrm{min}$ has been exaggerated to 0.5 deg/s for clarity. Not shown in the figure is a commandable overshoot $\delta \phi$ which the payload can travel beyond the scan endpoints without pcm switching over from the elliptical to the square root profile. }
\end{center}
\end{figure}

When developing the sinusoidal scan routine for pcm, the goal was to implement the scan described above in as stateless a way as possible. In order to determine $\omega_\mathrm{req}$ at any time, pcm needs to know only the current payload azimuth, and the direction of motion. The latter is measured by the rate gyroscopes, while the former is computed in the in-flight pointing solution~\cite{nng2014}. Since the velocity request is computed based on position, it is necessary to determine the velocity vs.~position profile for the sinusoidal scan. Taking $\omega(t) = \omega_\mathrm{req}$, Eqs.~\ref{eq:phi} and~\ref{eq:omega} can be combined to produce

\begin{equation}
\label{eq:ellipse}
\frac{(\phi-\phi_c)^2}{A^2} + \frac{\omega^2_\mathrm{req}}{\omega^2_\mathrm{pk}} = 1.
\end{equation}
The velocity vs.~position curve is an ellipse with semi-major and semi-minor axes given by $A$ and $\omega_\mathrm{pk}$. As shown in Figure~\ref{fig:ellipse}, the scan algorithm varies the velocity request based on payload azimuth as follows:
\\~\\~\\~\\
\begin{equation}
\label{eq:sinescan}
\omega_\mathrm{req} = \left\{ \begin{array}{lll}
         +\sqrt{2\alpha_\mathrm{pk}[(\phi_l - \delta \phi) - \phi]} + \omega_\mathrm{min}  & \mbox{if $\phi < (\phi_l - \delta \phi)$}; & \mbox{(beyond left scan endpoint)}\\~\\
         +\omega_\mathrm{min} & \mbox{if $\phi_l < \phi \leq (\phi_l + \Delta \phi_\mathrm{turn})$}; & \mbox{(in left turn-around zone)}\\~\\
         +\omega_\mathrm{pk}\sqrt{1 - (\phi - \phi_c)^2 / A^2} & \mbox{if $(\phi_l + \Delta \phi_\mathrm{turn}) < \phi < (\phi_r - \Delta \phi_\mathrm{turn})$} & \mbox{(in scan range and moving}\\
         & \mbox{and $\omega_\mathrm{gond} > 0$}; &\mbox{from left to right)}\\~\\
            -\omega_\mathrm{pk}\sqrt{1 - (\phi - \phi_c)^2 / A^2} & \mbox{if $(\phi_l + \Delta \phi_\mathrm{turn}) < \phi < (\phi_r - \Delta \phi_\mathrm{turn})$} & \mbox{(in scan range and moving}\\
         & \mbox{and $\omega_\mathrm{gond} < 0$}; &\mbox{from right to left)}\\~\\
         -\omega_\mathrm{min} & \mbox{if $(\phi_r - \Delta \phi_\mathrm{turn}) \leq \phi < \phi_r$}; & \mbox{(in right turn-around zone)}\\~\\
         -\sqrt{2\alpha_\mathrm{pk}[\phi - (\phi_r + \delta \phi)]} - \omega_\mathrm{min}  & \mbox{if $\phi > (\phi_r + \delta \phi)$}; & \mbox{(beyond right scan endpoint)}
         \end{array} \right.
\end{equation}
In Eq.~\ref{eq:sinescan}, $\phi_l \equiv \phi_c - A$ is the left az scan endpoint, and $\phi_r \equiv \phi_c + A$ is the right az scan endpoint. The parameter $\Delta \phi_\mathrm{turn}$ is the distance from the endpoints within which the speed on the elliptical profile drops below $\omega_\mathrm{min}$, prompting a turn-around. Finally, $\delta \phi$ is a permitted overshoot outside of the scan range.

\newpage
\subsection{The \textsc{\bfseries Spider} Observing Strategy}
\label{sec:spiderobs}

\begin{figure}[htbp!]
\begin{center}
\includegraphics[width=\textwidth]{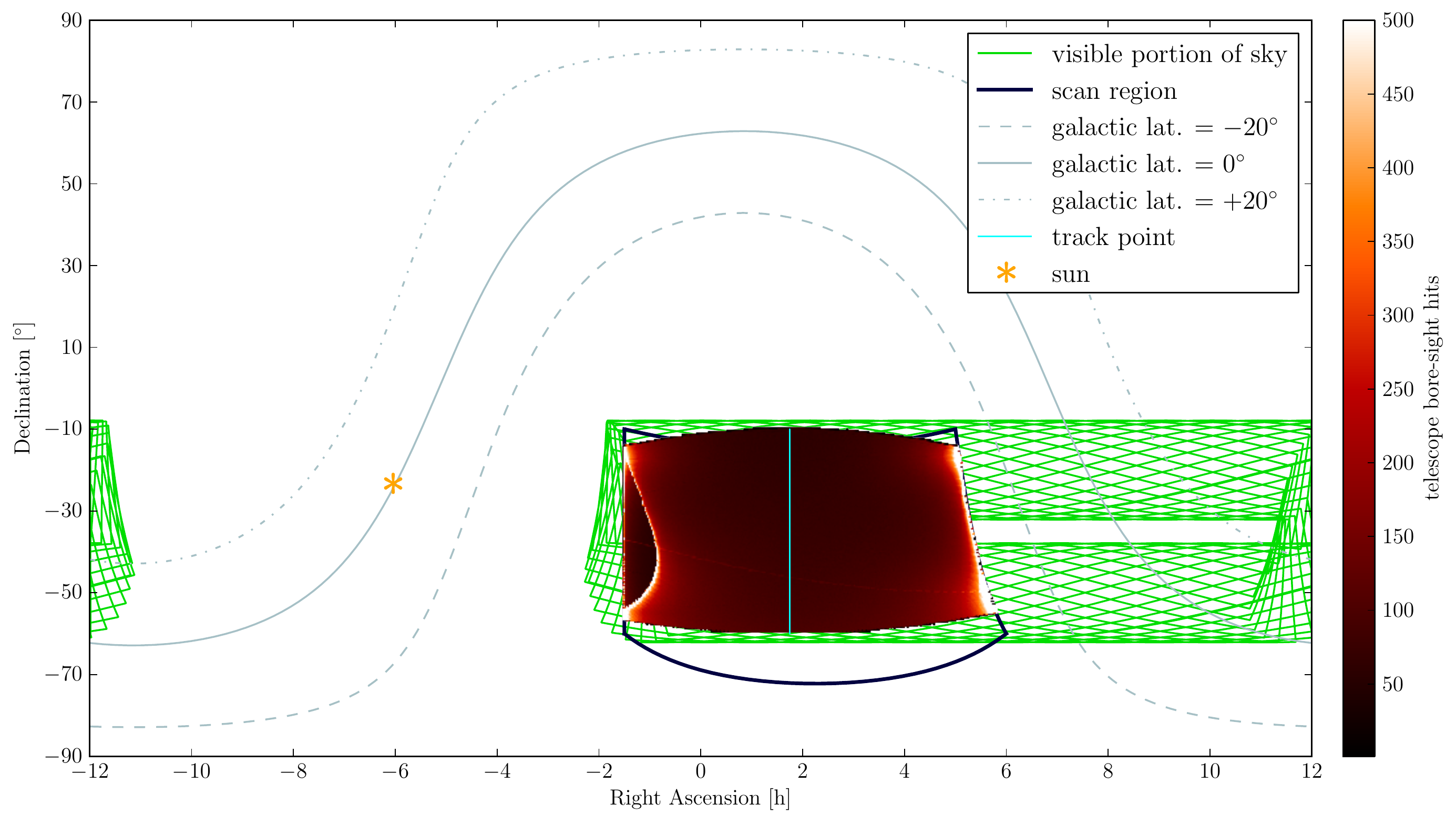}
\caption{\label{fig:scan_sim_fig} Output from a simulation of the \Spider~scan strategy spanning 24 solar hours. This simulation occurs at lat.~= 77.85$^\circ$ S, lon.~= 166.67$^\circ$ E, which are the coordinates of McMurdo Station, Antarctica. The dark blue outline is the defined scan region. The cyan vertical line traces out the path of the track point. Every hour, a green box has been drawn whose edges are \Spider's el limits and az limits relative to the sun. The union of these green boxes encompasses the total area of sky visible to \Spider, in principle, from this location. This simulation used el limits of $20^\circ \leq \theta \leq 50^\circ$. The az limit is $|\phi-\phi_\mathrm{sun}| \geq 70^\circ$ on the port side, and $|\phi-\phi_\mathrm{sun}| \geq 90^\circ$ on the starboard side The intensity map shows the number of hits for the \emph{telescope bore-sight} in pixels of Healpix $N_\mathrm{side}$ = 256, indicating the area of sky actually observed. The grey curves are lines of constant Galactic latitude.}
\end{center}
\end{figure}

The scan strategy to be used for CMB observations has the sinusoidal az scan from Section~\ref{sec:sinescan} as a basis, but includes the following additional elements. A quadrangular region (hereafter simply ``the box''), within which observations are to be confined, is defined on the sky. This box is the thick blue outline in Figure~\ref{fig:scan_sim_fig}. The coordinates of the four corner points of the box in right ascension (RA) and declination (dec) are specified. These corner points are connected by great-circle arcs. This method makes it easy to avoid observing too close to the Galactic plane; the defined scan region simply does not encompass these areas. The box also determines sky coverage, subject to additional constraints from the sun azimuth and mechanical elevation limits (Figure~\ref{fig:scan_sim_fig}). 

A specific point within the box, known as the ``track point'', is chosen in RA and dec. The instantaneous sinusoidal az scan is constrained to pass through this point, and to terminate on the edges of the box. Therefore, at the BLASTbus frame rate of $\sim$120~Hz, pcm transforms from equatorial to horizon coordinates to determine the track point elevation, and the scan endpoints in azimuth $\phi_r$ and $\phi_l$ of a line at this elevation that is confined to the box. The requested elevation angle $\theta_\mathrm{req}$ (Section~\ref{sec:elctrl}) is set to the computed track point elevation. The results of these coordinate transformations will change with latitude and with local sidereal time (LST). Therefore, the science az scan is like the one described in Section~\ref{sec:sinescan}, but with a time-variable $\phi_r$ and $\phi_l$, and hence a time-variable $A$ and $\phi_c$.

The RA of the track point is kept constant, while its dec is varied back and forth between the top and bottom edges of the box at a constant rate with LST. It reaches its highest declination when the hour angle (LST $-$ RA) of the track point is zero, meaning that the centre of the box reaches its highest point, crossing the meridian. The declination of the track point reaches its lowest value 12 sidereal hours later. This phase of the dec vs.~LST variation maximizes sky coverage. The purpose of moving the track point is to ensure even coverage of the box in dec, filling in the gaps between rows of detector beams on the sky. Rather than drifting continuously, \Spider's elevation angle is updated at every $n^\mathrm{th}$ scan turn-around in order to keep up with the motion of the track point. The parameter $n$ is an integer that can be commanded in flight.
\begin{table}[htbp!]
\caption{A strategy for stepping the \Spider~HWPs } 
\label{tab:hwp}
\begin{center}       
\begin{tabular}{cr|c|cr}
\hline
\rule[-1ex]{0pt}{3.5ex} \textbf{Sidereal Day} & \textbf{HWP Angle} & & \textbf{Sidereal Day} & \textbf{HWP Angle}  \\
\hline 
\hline 
\rule[-1ex]{0pt}{3.5ex}  1.0 & 0.0$^\circ$ & & 5.0 & 90.0$^\circ$   \\
\rule[-1ex]{0pt}{3.5ex}  1.5 & 22.5$^\circ$ & & 5.5 & 112.5$^\circ$   \\
\rule[-1ex]{0pt}{3.5ex}  2.0 & 45.0$^\circ$ & & 6.0 & 135.0$^\circ$  \\
\rule[-1ex]{0pt}{3.5ex}  2.5 & 67.5$^\circ$ & & 6.5 & 157.5$^\circ$   \\
\rule[-1ex]{0pt}{3.5ex}  3.0 & 22.5$^\circ$ & & 7.0 & 112.5$^\circ$   \\
\rule[-1ex]{0pt}{3.5ex}  3.5 & 45.0$^\circ$ & & 7.5 & 135.0$^\circ$   \\
\rule[-1ex]{0pt}{3.5ex}  4.0 & 67.5$^\circ$ & & 8.0 & 157.5$^\circ$   \\
\rule[-1ex]{0pt}{3.5ex}  4.5 & 90.0$^\circ$ & & 8.5 & 0.0$^\circ$   \\
\hline
\end{tabular}
\end{center}
\end{table} 

A preliminary strategy for HWP stepping is given in Table~\ref{tab:hwp}. In this table, integer sidereal days begin when the bore-sight is at the bottom of the box, moving upwards in dec. Half-integer sidereal days begin once the scan has reached the top of the box and begins moving back downward. Therefore, a bolometer whose polarization axis is oriented to measure $+Q$ will do so during the up-going scan in the first half of day 1.0. The $22.5^\circ$ HWP step will then switch its sensitivity to $+U$ on the down-going scan. The same bolometer will undergo the sequence $-Q$ and $-U$ on sidereal day 2.0. Therefore, after two sidereal days, each bolometer should have made an independent measurement of the Stokes polarization vector in each sky pixel. In sidereal days 3.0 and 4.0, the same sequence of HWP steps occurs, but shifted by +22.5$^\circ$, resulting in the sequence ($+U, -Q, -U, +Q$) for a $+Q$-oriented bolometer. In sidereal days 5-8, the HWP angles are shifted by $+90^\circ$ from the sequence in days 1-4. This method has the advantage that, after 8 sidereal days, both up-going and down-going scans of the box exist for \emph{each} unique HWP position.

\section{SYSTEM PERFORMANCE}
\label{sec:sys}

\subsection{Azimuth Velocity Control}
\label{sec:v_az_sys}
\begin{figure}[htbp!]
\begin{center}
\includegraphics[width=0.88\textwidth]{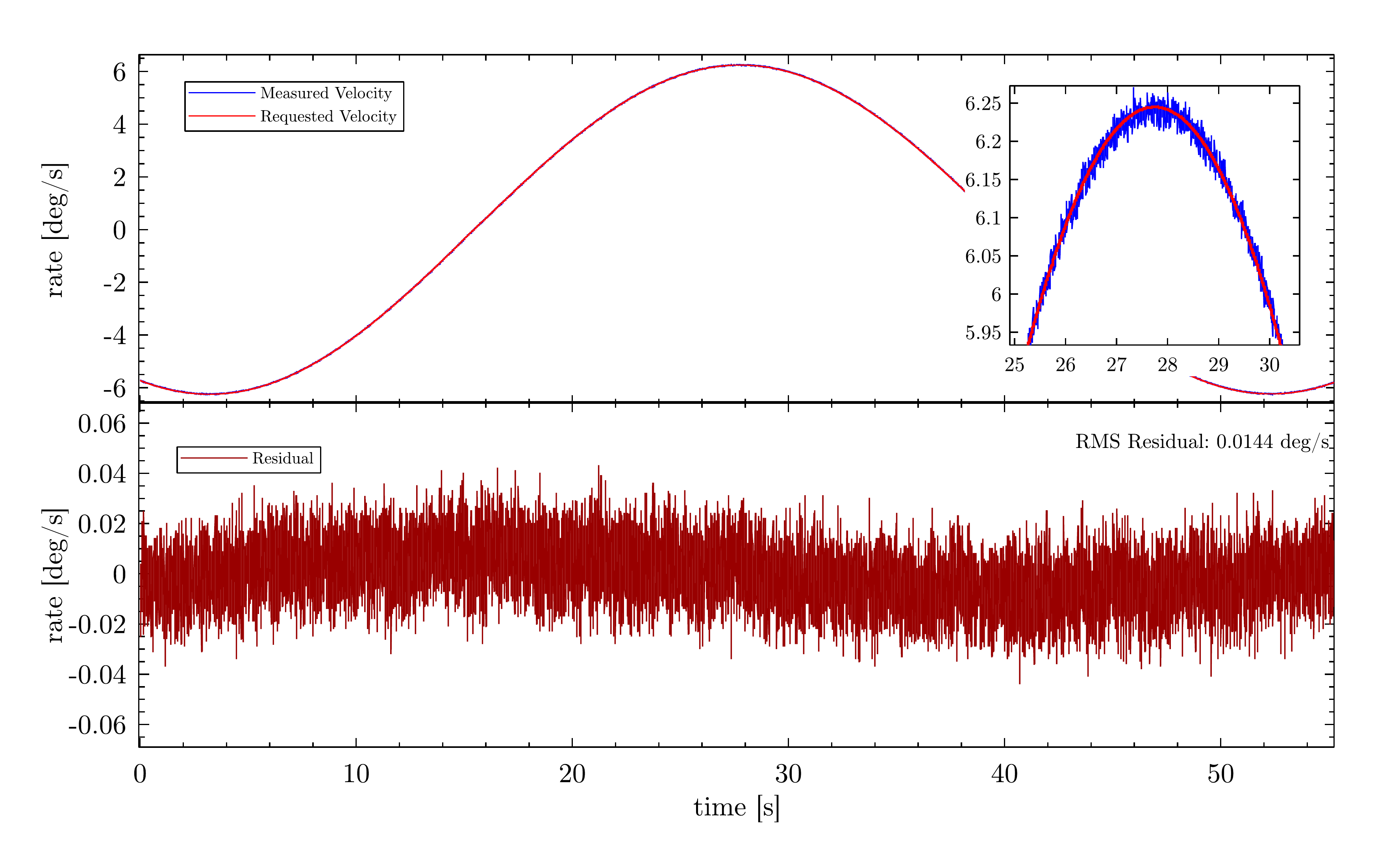}
\caption{\label{fig:vel_sys} \textit{Top} -- Requested and measured angular velocities vs.~time during approximately one scan period. \textit{Inset} -- A close up near the peak of the sinusoidal profile (i.e.~at the scan centre). \textit{Bottom} -- The velocity error, with the RMS value indicated. } 
\end{center}
\end{figure}

Figure~\ref{fig:vel_sys} indicates how well the azimuth velocity control described in Section~\ref{sec:az} performs. These data were obtained during scan tests conducted at CSBF on August 8, 2013. The payload was almost fully-assembled, as depicted in Figure~\ref{fig:spider} (\textit{right}). It was suspended from a hoist inside one of the CSBF high bays. As shown in the figure, the payload angular velocity, measured by the yaw-axis gyros, tracks the requested sinusoidal profile with high fidelity. In addition to actual pointing control error, gyro noise contributes to the RMS residual. The RMS noise level in the gyro timestreams is measured to be about half the RMS residual given in Figure~\ref{fig:vel_sys}. 

\subsection{Elevation Position Control}

\begin{figure}[htbp!]
\begin{center}
\includegraphics[width=0.88\textwidth]{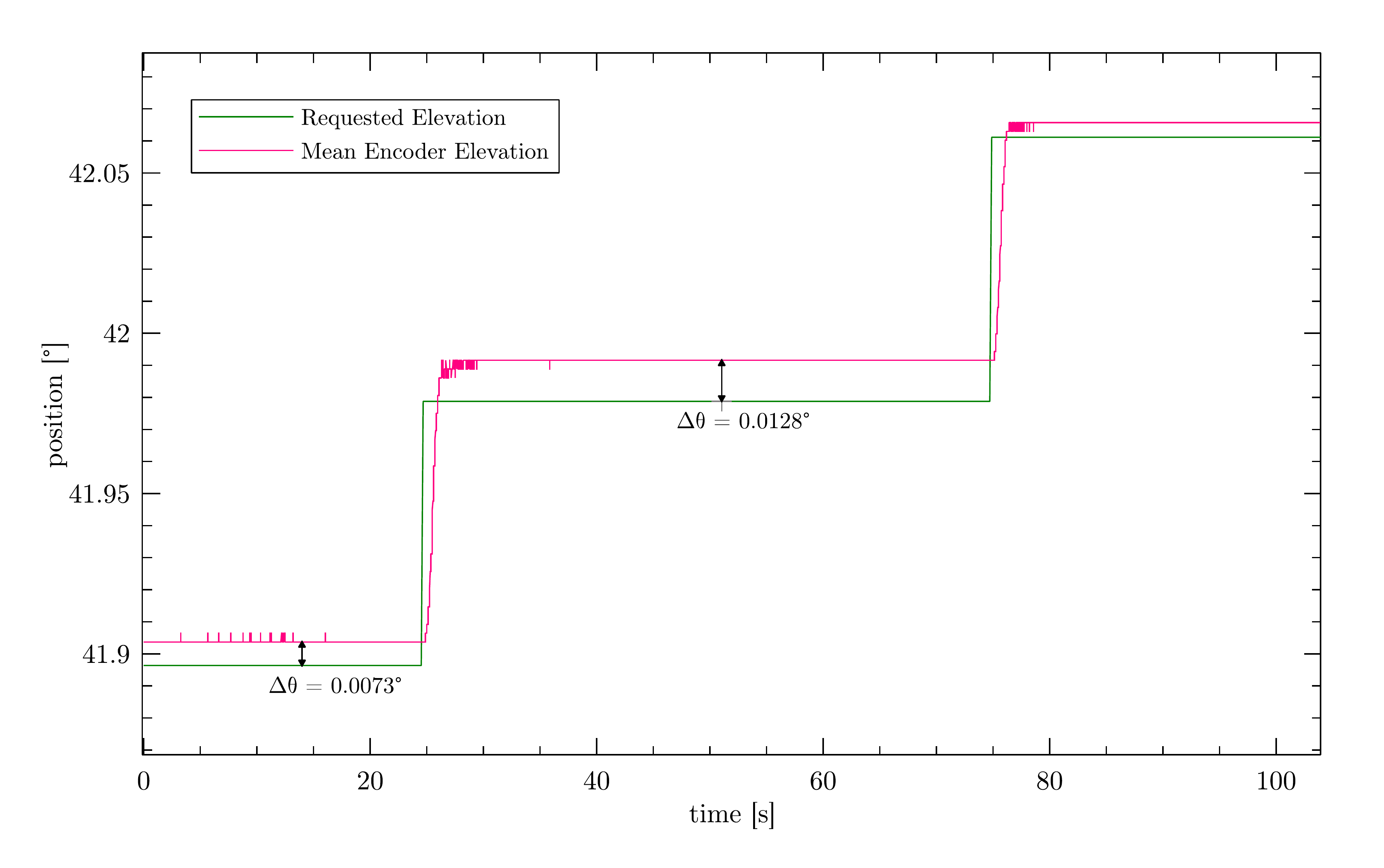}
\caption{\label{fig:el_sys} Requested and measured elevation angles for two el steps at consecutive scan turn-arounds during a test of the science scan mode. The position error $\Delta\theta$ is shown before and after the first step. Both values are within pcm's tolerance of 0.02$^\circ$.  } 
\end{center}
\end{figure}

During the scan tests described in the previous section, the elevation drive was consistently able to servo el position to within 4 encoder counts or 0.02$^\circ$ (Figure~\ref{fig:el_sys}), which is the position error tolerance allowed by pcm. 


\newpage
\subsection{Testing The Scan Strategy}
\begin{figure}[hbtp!]
\begin{center}
\includegraphics[width=\textwidth]{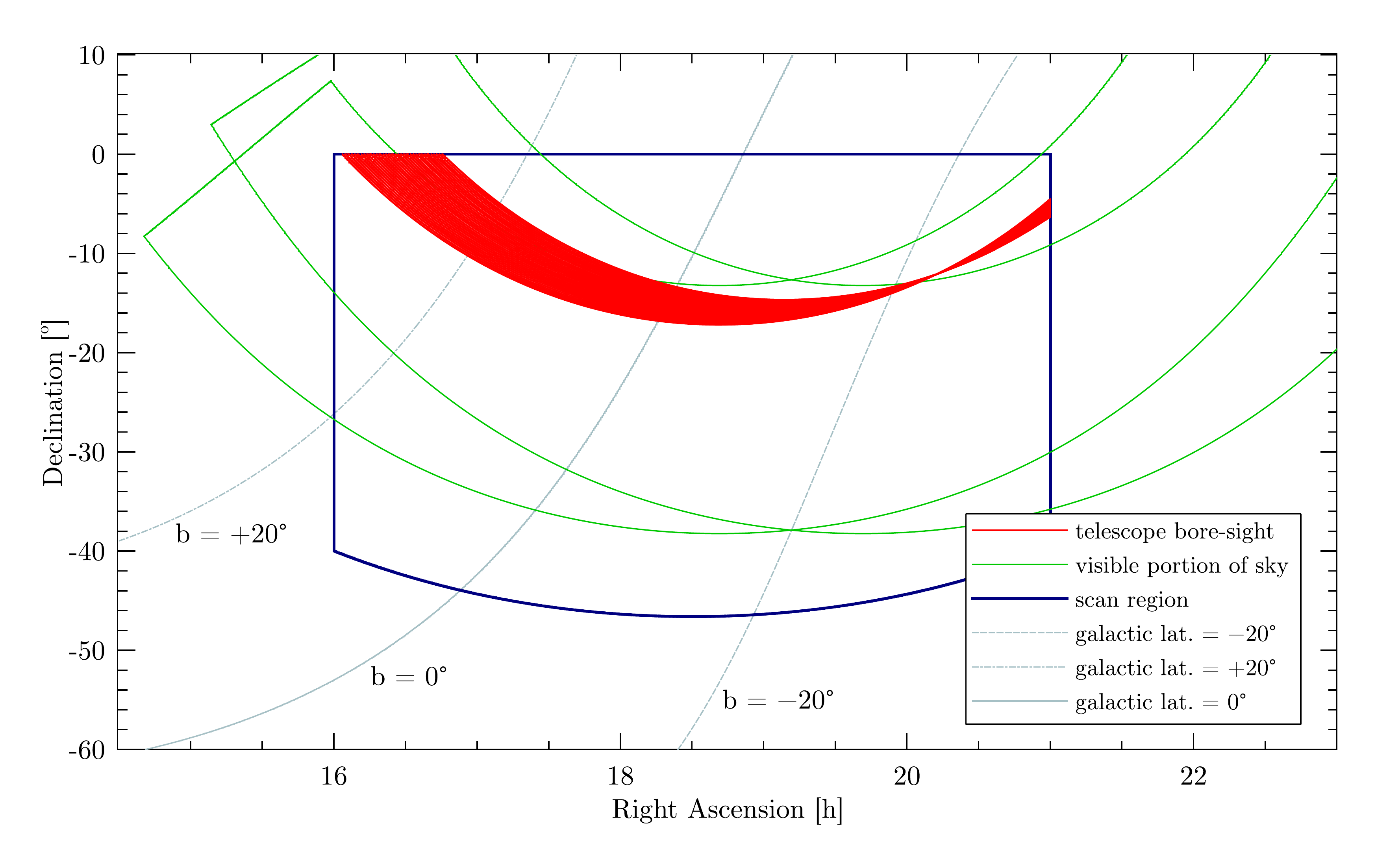}
\caption{\label{fig:scan_test} A plot of a science-mode \Spider~scan in equatorial coordinates. The red curve traces the position of the telescope bore-sight on the sky with time. These are actual data from the \Spider~in-flight pointing solution spanning approximately 30 minutes of scan testing, starting at 22:52:46 CDT on August 8, 2013. The scan tests took place in Palestine, TX, USA (31.7619$^\circ$ N, 95.6306$^\circ$ W), which explains the area of sky that was observed. For flight, \Spider~will launch from Antarctica and scan over a region low in foreground emission, far from the Galactic plane.} 
\end{center}
\end{figure}

During the scan testing referred to in Section~\ref{sec:v_az_sys}, pcm was computing the in-flight pointing solution. As a result, the approximate position of the telescope bore-sight vs.~time was recorded in the downlink data at the BLASTbus frame rate of approximately 120~Hz. These data are shown in Figure~\ref{fig:scan_test}, which indicates that the algorithm of Section~\ref{sec:obs} works as intended. Although higher-level details of the scan strategy have changed since these tests were performed, the base scan algorithm was exactly as described in Section~\ref{sec:sinescan}.

\section{CONCLUSIONS}
\label{sec:end}

The pointing control system that has been developed meets the observing requirements of a balloon-borne CMB polarimeter 
aiming to map approximately 10\% of the sky at degree angular scales. Some of the hardware and control software is derived from that of past experiments, particularly BLASTPol. However, significant innovations were required in order to produce a control system suitable for \Spider. Some of these include:

\begin{itemize}

\item The design, integration and testing of the elevation drive hardware. The elevation drive system had to be able to support and torque on the massive flight cryostat, and it consists of components that have not flown on any previous balloon payload.
\item The development of a position control algorithm for the elevation drive. This task posed a challenge due to the presence of static friction, phase lag, and an unbalanced load in the system. 
\item The development of an algorithm for operating the pivot in velocity control mode, to avoid driving pendulations of the gondola. This was especially important given the large amplitude and angular acceleration of \Spider~scans.
\item The development of the \Spider~observing strategy. The azimuth scan is fast enough and wide enough to overcome $1/f$ noise and to ensure sufficient coverage, respectively. The strategy of tracking a moving point on the sky with LST ensures even coverage in declination, with no gaps.

\end{itemize}

\noindent Laboratory testing to date has shown a performance of the control system well within acceptable limits, although the system has not yet been used in a balloon flight. \Spider's first Long Duration Balloon flight is scheduled for December 2014.

\acknowledgments     

The \Spider\ collaboration gratefully acknowledges the support of NASA (award numbers NNX07AL64G, NNX12AE95G), the Lucille and David Packard Foundation, the Gordon and Betty Moore Foundation, the Natural Sciences and Engineering Research Council (NSERC), the Canadian Space Agency (CSA), and the Canada Foundation for Innovation. We thank the JPL Research and Technology Development Fund for advancing detector focal plane technology.  W.\,C.\,Jones acknowledges the support of the Alfred P. Sloan Foundation. A.\,S.\,Rahlin is partially supported through NASAs NESSF Program (12-ASTRO12R-004). J.\,D.\,Soler acknowledges the support of the European Research Council under the European Union's Seventh Framework Programme FP7/2007-2013/ERC grant agreement number 267934.

Logistical support for this project in Antarctica is provided by the U.S. National Science Foundation through the U.S. Antarctic Program. We would also like to thank the Columbia Scientific Balloon Facility (CSBF) staff for their continued outstanding work.


\bibliography{jas_spider_spie}   
\bibliographystyle{spiebib}   

\end{document}